\documentclass[twocolumn]{aastex63}
\usepackage{lineno}
\begin{document}

\title{The Hubble PanCET program: Transit and Eclipse Spectroscopy of the Hot Jupiter WASP-74b}

\author[0000-0002-3263-2251]{Guangwei Fu}
\author{Drake Deming}
\affiliation{Department of Astronomy, University of Maryland, College Park, MD 20742, USA; guangweifu@gmail.com}

\author{Erin May}
\author{Kevin Stevenson}
\affiliation{Johns Hopkins APL, 11100 Johns Hopkins Rd, Laurel, MD 20723, USA}

\author{David K. Sing}
\author{Joshua D. Lothringer}
\affiliation{Department of Physics and Astronomy, Johns Hopkins University, Baltimore, MD 21218, USA}

\author[0000-0003-4328-3867]{H.R. Wakeford}
\affil{School of Physics, University of Bristol, HH Wills Physics Laboratory, Tyndall Avenue, Bristol BS8 1TL, UK}

\author{Nikolay Nikolov}
\affiliation{Space Telescope Science Institute, 3700 San Martin Dr, Baltimore, MD 21218, USA}

\author[0000-0001-5442-1300]{Thomas Mikal-Evans}
\affiliation{Department of Physics, and Kavli Institute for Astrophysics and Space Research, Massachusetts Institute of Technology, Cambridge, USA}

\author{Vincent Bourrier}
\author[0000-0002-2248-3838]{Leonardo A. dos Santos}
\affil{Observatoire astronomique de l'Universit\`e de Gen\`eve, 51 chemin des Maillettes 1290 Versoix, Switzerland}

\author[0000-0003-4157-832X]{Munazza K. Alam}
\affiliation{Department of Astronomy, Center for Astrophysics $|$ Harvard \& Smithsonian, 60 Garden Street, Cambridge, MA 02138, USA}
\affiliation{Carnegie Earth \& Planets Laboratory, 5241 Broad Branch Rd NW, Washington, DC 20015, USA}

\author{Gregory W. Henry}
\affiliation{Tennessee State University, Center of Excellence in Information Systems, Nashville, TN  37209, USA}

\author{Antonio Garc\'ia Mu\~noz}
\affiliation{AIM, CEA, CNRS, Universit\'e Paris-Saclay, Universit\'e de Paris, F-91191 Gif-sur-Yvette, France}

\author[0000-0003-3204-8183]{Mercedes L\'{o}pez-Morales}
\affiliation{Center for Astrophysics ${\rm \mid}$ Harvard {\rm \&} Smithsonian, 60 Garden Street, Cambridge, MA 02138, USA}

\begin{abstract}

Planets are like children with each one being unique and special. A better understanding of their collective properties requires a deeper understanding of each planet. Here we add the transit and eclipse spectra of hot Jupiter WASP-74b into the ever growing dataset of exoplanet atmosphere spectral library. With six transits and three eclipses using the Hubble Space Telescope (HST) and Spitzer Space Telescope (\textit{Spitzer}), we present the most complete and precise atmospheric spectra of WASP-74b. We found no evidence for TiO/VO nor super-Rayleigh scattering reported in previous studies. The transit shows a muted water feature with strong Rayleigh scattering extending into the infrared. The eclipse shows a featureless blackbody-like WFC3/G141 spectrum and a weak methane absorption feature in the Spitzer 3.6 $\mu m$ band. Future James Webb Space Telescope (JWST) follow up observations are needed to confirm these results.

\end{abstract}
\nopagebreak

\section{Introduction}

With the many recent studies of hot Jupiter atmospheres, we are starting to see a highly diverse landscape of atmospheric chemical composition and thermal structure. Some ultra-hot (T$_{eq}>$2000K) Jupiters have shown significant gaseous heavy metal absorption \citep{fu_hubble_2021, lothringer_uv_2020} while other cooler hot Jupiters exhibit high metallicity \citep{colon_unusual_2020, sheppard_hubble_2021, lewis_into_2020} and aerosols coverage \citep{alam_hubble_2020}. We have also detected absorption and emission features of various chemical species such as water, sodium, CO/CO$_2$, metals and metal oxides (TiO and VO) \citep{evans_ultrahot_2017, fu_hubble_2021, kreidberg_water_2018}. With this ever growing library of exoplanet atmosphere spectra, we are starting to understand the chemical and physical processes taking place in these hot Jupiters. For example, statistical studies \citep{fu_statistical_2017, sing_continuum_2016} have provided valuable insights into how aerosol presence varies in different atmospheric conditions and were later supported by detailed aerosol modeling \citep{gao_aerosol_2020}. Comparative studies \citep{parmentier_thermal_2018, garhart_statistical_2020, zhang_phase_2018, baxter_evidence_2021, wallack_trends_2021} have furthered our understanding of exoplanet thermal structures and heat circulation \citep{kataria_atmospheric_2016}. A large library of measured high quality high precision exoplanetary spectra will be essential to our future studies of exoplanet atmospheres.

Here we present an uniformly analyzed transmission and emission spectrum of hot Jupiter WASP-74b (Table \ref{W74}) ranging from 0.3 to 4.5 $\mu m$ using multiple instruments on the Hubble Space Telescope (HST) and Spitzer Space Telescope (\textit{Spitzer}). We have also compared our transit spectrum with previous ground based observations and other independent analyses. We found a muted water absorption feature with a significant aerosol scattering slope in the transit spectrum, and a featureless blackbody-like WFC3/G141 eclipse spectrum with absorption feature in the Spitzer 3.6 $\mu m$ bands.

\section{Observations}

We observed a total of six transits and three eclipses of WASP-74b using multiple instrument modes on HST and Spitzer (Table. \ref{observations}). The data comprises two observation programs and some parts of the data were published in previous studies \citep{luque_obliquity_2020, mancini_physical_2019, garhart_statistical_2020}. We have uniformly analyzed the full dataset including new HST/STIS transit, HST/WFC3 and Spitzer eclipse. All the data are reduced using the same orbital parameters and limb darkening coefficients from the same 3D stellar model.  

\begin{table*}[t]
\centering
\begin{tabular}{cccccc}
\multicolumn{6}{c}{\textbf{WASP-74b transit observations}}\\
\hline\hline 
&	Grism/Filter	&	Visit 1	&	Visit 2	   &   GO Program ID & PI \\   
\hline 
HST STIS	&	G430L	&	2017-05-04	&	  2017-07-20	&	14767	& Sing $\&$ L\'{o}pez-Morales \\
HST STIS	&	G750L	&	2017-06-20	&		&	14767	&	Sing $\&$ L\'{o}pez-Morales	\\
HST WFC3	&	G141	&	2016-10-06	&		&	14767	&	Sing $\&$ L\'{o}pez-Morales	\\
Spitzer	&	IRAC 3.6	&	2017-01-14	&	 	&	13044	&	Deming	\\
Spitzer	&	IRAC 4.5	&	2017-01-16	&		&	13044	&	Deming	\\ 
\hline 
\smallskip \\
\multicolumn{6}{c}{\textbf{WASP-74b eclipse observations}}\\
\hline\hline
&	Grism/Filter	&	Visit 1   &   GO Program ID & PI \\   
\hline 
HST WFC3	&	G141	&	2017-05-02	&	14767	&	 Sing $\&$ L\'{o}pez-Morales	\\
Spitzer	&	IRAC 3.6	&	2017-01-15	&	13044	&	Deming	\\
Spitzer	&	IRAC 4.5	&	2017-02-12	&	13044	&	Deming	\\
\hline 

\end{tabular}
\caption{A list of our 6 transit and 3 eclipse observations of WASP-74b.}
\label{observations}
\end{table*}

\begin{center}
\begin{table}[]
\hskip-1.0cm\begin{tabular}{ccc}
\multicolumn{3}{c}{\textbf{WASP-74b parameters}} \\
\hline\hline
Equilibrium Temperature	&	T$_{eq}$ (K)	&	$1926^{+21}_{-21}$	\\
Radius	&	R$_p$ (R$_{Jup}$)	&	$1.404^{+0.018}_{-0.012}$	\\
Mass	&	M$_p$ (M$_{Jup}$)	&	$0.826^{+0.015}_{-0.014}$	\\
Period	&	P$_{orb}$ (days)	&	$2.1377445^{+0.0000018}_{-0.0000018}$	\\
Semi-major axis	&	a (au)	&	$0.03443^{+0.00022}_{-0.00029}$	\\
Inclination	&	i (degree)	&	$79.86^{+0.21}_{-0.21}$	\\
\hline 
\end{tabular}
\caption{Parameters are from \citep{mancini_physical_2019}.}
\end{table}
\label{W74}
\end{center}

\section{Data analysis}

\subsection{HST STIS data reduction}

HST orbits around the earth every 90 minutes which results in an observing gap of $\sim$45 minutes when the target is not in the continuous viewing zone. Combined with the thermal breathing of HST, this leads to an orbital-dependent systematics of the lightcurve. Various detrending techniques \citep{sing_hubble_2011, nikolov_hubble_2014} have been developed to correct for this effect. \cite{sing_hubble_2019} used the time-tagged engineering information from the jitter files for each exposure to construct a parametric model. This new jitter decorrelation method was shown to improve the photometric performance of HST STIS time series observations compared to previous used methods \citep{sing_hubble_2019}. 

The STIS data reduction pipeline we used in this paper is identical to the jitter decorrelation method detailed in \cite{sing_hubble_2019} except it is applied on G430L and G750L data instead of NUV-MAMA detector with E230M echelle grating. The orbital parameters (Table. \ref{W74}) have been fixed to the values used in \cite{mancini_physical_2019}. For limb darkening (Table \ref{LD}), we used the 3D stellar model from
the Stagger-grid \citep{magic_stagger-grid_2015} with the parameters $T_{eff}$=6000K, log $g$=4, [Fe/H]=0. The systematics corrected lightcurves and residuals for two G430L visits and one G750L visit are shown in Figures \ref{fig:STIS_v76}, \ref{fig:STIS_v77} and \ref{fig:STIS_v78}. 

\subsection{HST WFC3 data reduction}

The HST WFC3 data reduction pipeline is exactly the same as detailed in \cite{fu_hubble_2021} for both transit and eclipse. We used a combination of the \texttt{BATMAN} lightcurve model \citep{kreidberg_batman_2015} with the \texttt{RECTE} charge trapping systematics model \citep{zhou_physical_2017} for lightcurve detrending. The orbital parameters (Table. \ref{W74}) have been fixed to the values used in \cite{mancini_physical_2019} and the limb darkening coefficients are from the same stellar model used for the STIS data reduction.

\subsection{Spitzer data reduction}

Two transits and eclipses of WASP-74b were observed with Spitzer's InfraRed Array Camera \citep[IRAC][]{fazio_infrared_2004}, at both 3.6 $\micron$ and 4.5 $\micron$ (PI: D. Deming, Program ID 13044). Table \ref{observations} includes the observation dates for these events. Each event used an exposure time of 0.4 seconds and covered approximately 7 hours total. For data reduction and analysis we use the Photometry for Orbits, Eclipses, and Transits pipeline \citep[POET][]{campo_orbit_2011, stevenson_transit_2012, cubillos_wasp-8b_2013}. 

\begin{figure*}
    \centering
    \includegraphics[width = 0.24\textwidth]{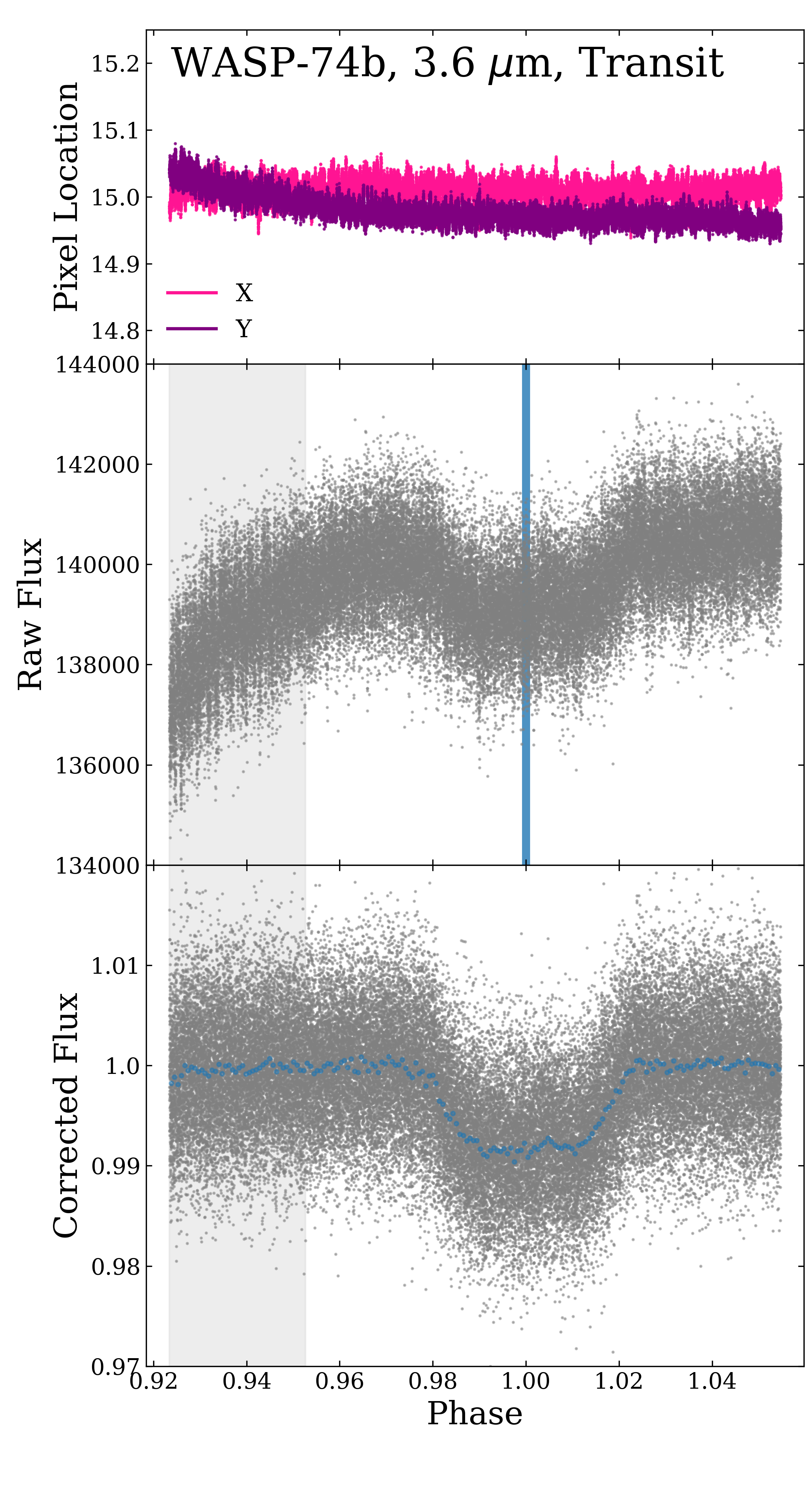}
    \includegraphics[width = 0.24\textwidth]{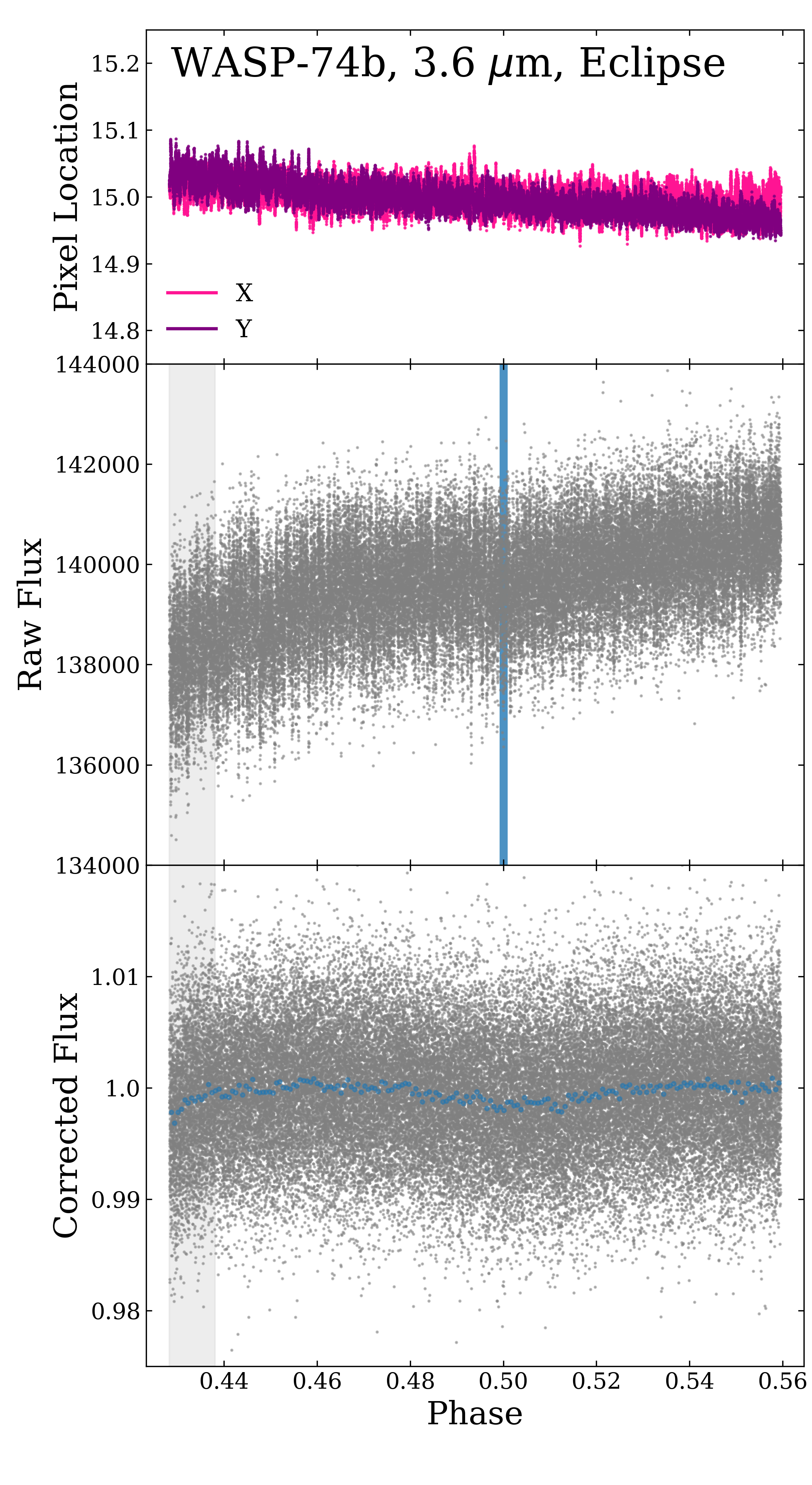}
    \includegraphics[width = 0.24\textwidth]{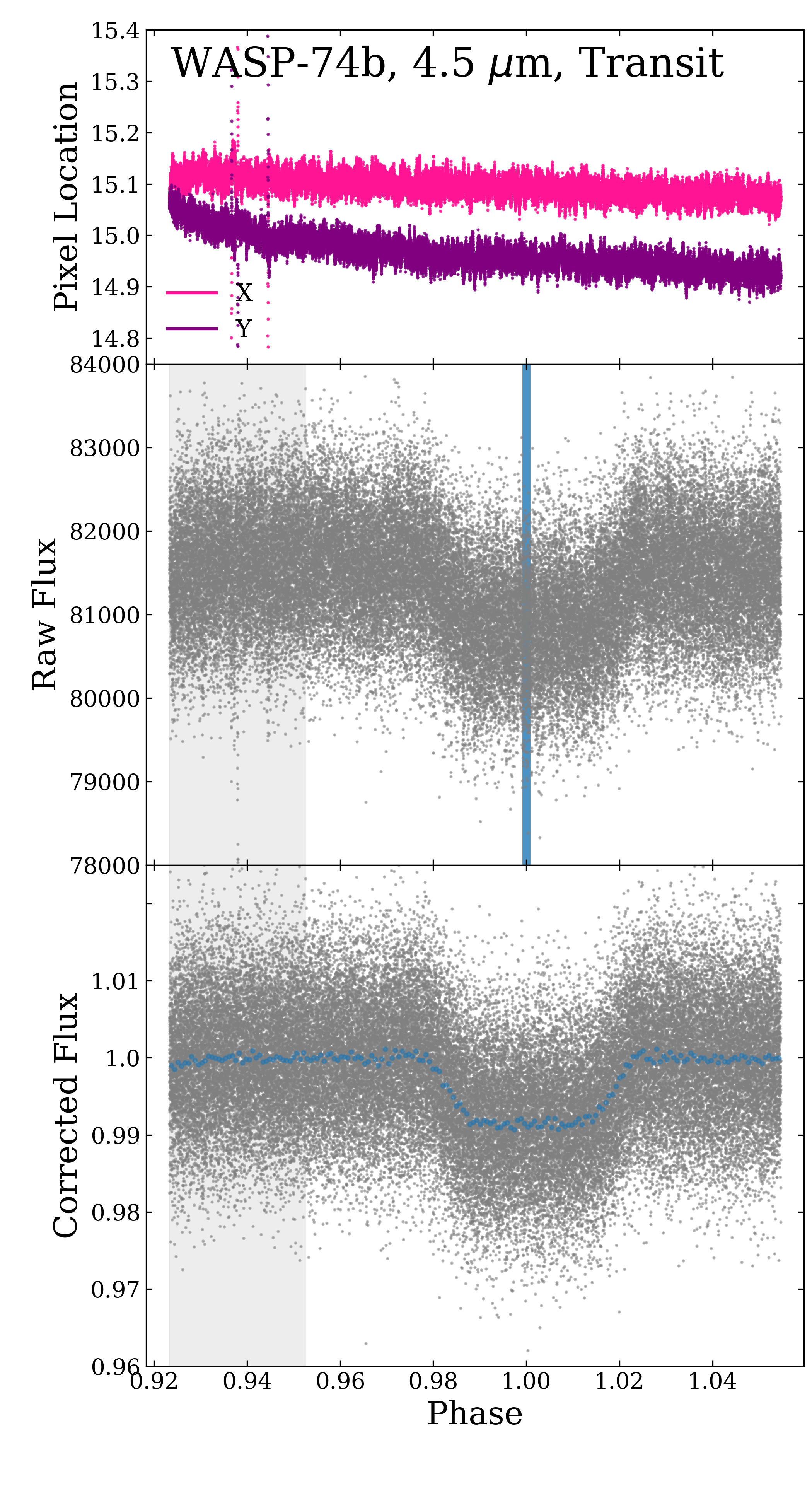}
    \includegraphics[width = 0.24\textwidth]{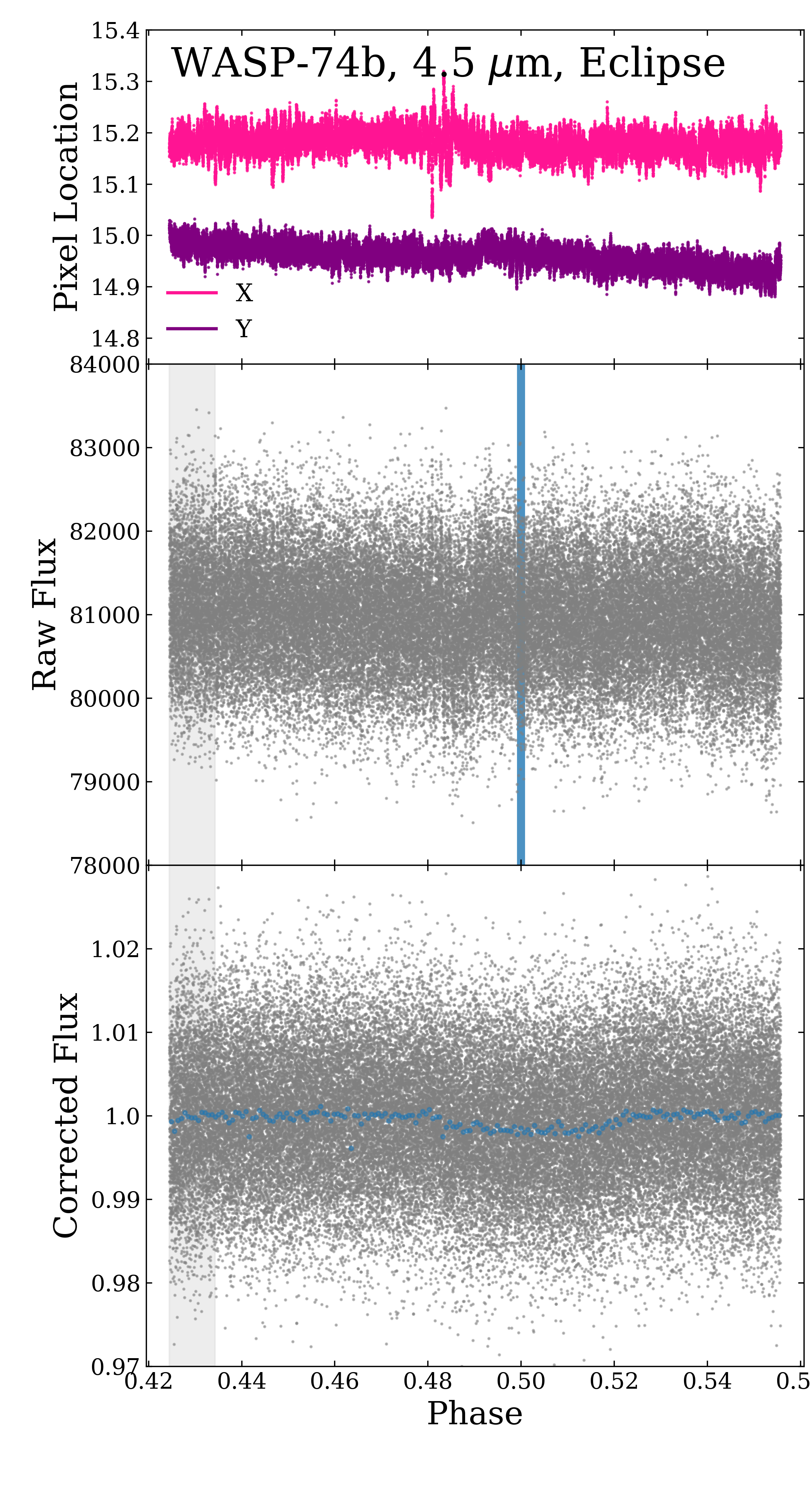}
    \caption{Centroids (top panels), raw flux (middle panels) and corrected flux (bottom panels) as a function of phase for all four Spitzer IRAC data sets as labeled. One can see that the 3.6 $\micron$ data sets are more heavily affected by a visit-long ramp, in combination with a stronger initial ramp for the 3.6 $\micron$ transit event. The trimmed parts of the lightcurves are shaded in grey.}
    \label{fig:centroids}
\end{figure*}

Spitzer IRAC photometry has been the major instrument mode used to investigate exoplanet atmospheres in the infrared \citep{deming_highlights_2020}. The raw data is usually dominated by instrumental systematics due to intrapixel sensitivity variations on the
detector in both 3.6 and 4.5 $\mu m$ channels \citep{deming_spitzer_2015}.
During the past decade we have developed multiple methods \citep{deming_spitzer_2015, stevenson_transit_2012} to mitigate this effect. Recently, \citep{may_introducing_2020, may_spitzer_2021} generated a new Gaussian centroided intrapixel sensitivity map based on years of archival Spitzer calibration data. Here we use this fixed sensitivity map \cite{may_introducing_2020} to analyze the 4.5 $\micron$ Spitzer datasets while using standard BLISS Mapping at 3.6 $\micron$. At 3.6 $\micron$ we also find that the inclusion of a second-order function with respect to the widths derived from 2D Gaussian fits (PRF-FWHM) is optimal. Figure \ref{fig:centroids} shows the x and y pixel locations and raw flux for all Spitzer IRAC data sets analyzed here.

For the astrophysical signal, we fit transits using the \texttt{BATMAN} \citep{kreidberg_batman_2015} with limb darkening coefficients fixed to values taken from the same stellar models as those used for the optical data reductions. Eclipses are modeled using the analytical model in \cite{mandel_analytic_2002}. All orbital parameters (Table. \ref{W74}) are fixed to the same values used in the HST STIS and WFC3 data reduction to minimize systematic offsets between the data sets.

\begin{figure*}
\centering
  \includegraphics[width=\textwidth,keepaspectratio]{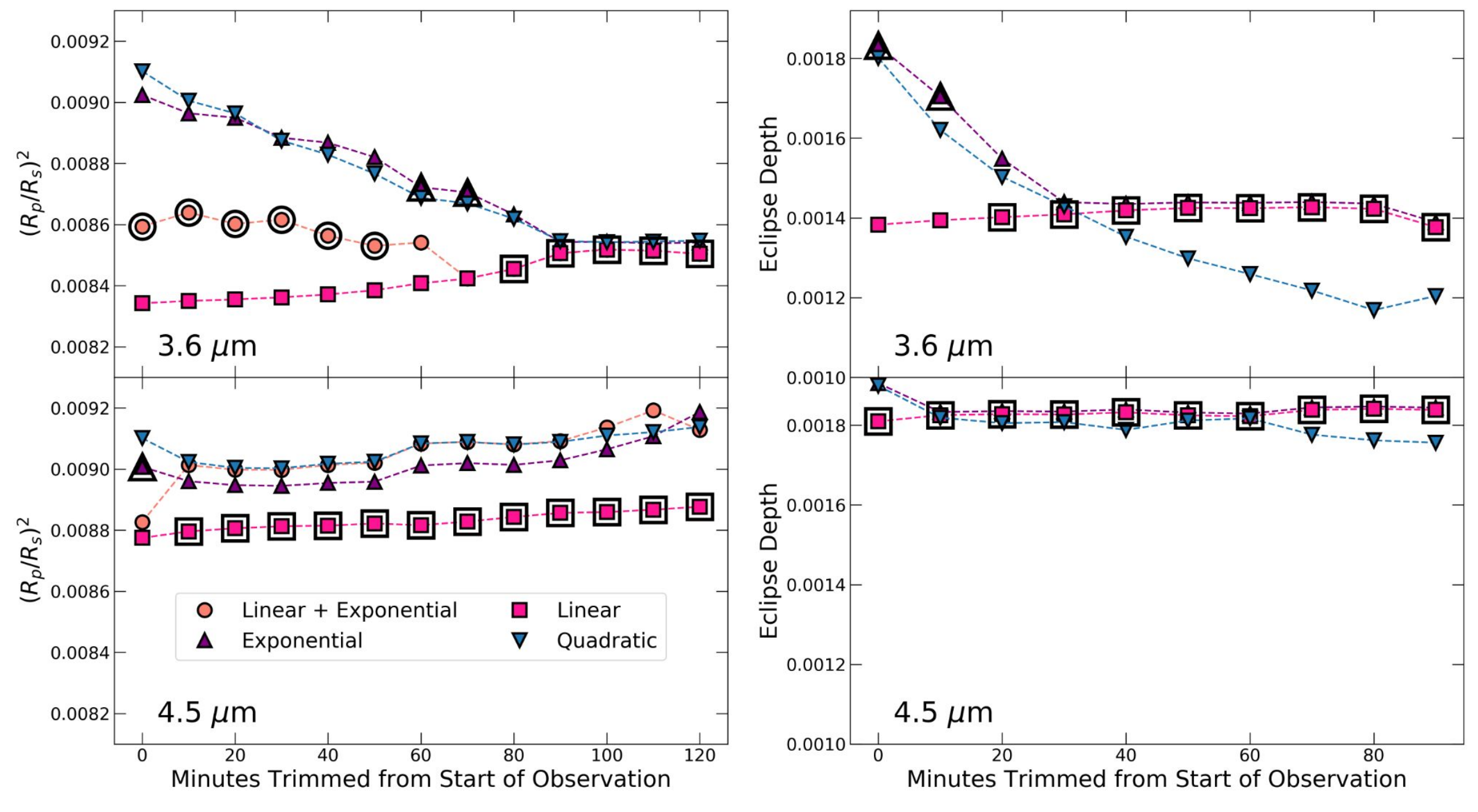}
  \caption{\textbf{Demonstration of the dependence on ramp choice and amount of data trimmed from the start of events}. The 3.6 $\micron$ and 4.5 $\micron$ events are shown on the same scale for ease of comparison. Left: Transits, Right: Eclipses. Top: 3.6 $\micron$, Bottom: 4.5 $\micron$. Our use of the fixed sensitivity map for systematic removal at 4.5 $\micron$ results in a consistent eclipse and transit depth regardless of pre-trimming and ramp model, while the 3.6 $\micron$ events are heavily dependent on the choice of these parameters due to the flexibility of the BLISS map. }
  \label{fig:Spitzer}
\end{figure*}

At the beginning of the observation, instrument systematics are the strongest due to the time needed for the telescope pointing to stabilize and the detector to fill up the charge traps. Therefore, the resulting ramp effect at the beginning is challenging to model and the first portion of the data has usually been discarded \citep{stevenson_transit_2012, garhart_statistical_2020, baxter_evidence_2021}. For this work, we have experimented varying the amount of data points to trim at the beginning of the observation from 0 to 120 minutes while fitting the remaining lightcurve with four different ramp models (Fig. \ref{fig:Spitzer}, left panels show transits and right panels show eclipses). 

As expected, because the fixed sensitivity map is not dependent on other modeling choices like a standard BLISS map, its use at 4.5 $\micron$ results in minimal variations of the measured eclipse and transit depths with the amount of data trimmed from the beginning (bottom two panels of Figure \ref{fig:Spitzer}), while the free BLISS map used for 3.6 $\micron$ results in transit and eclipse depths which are heavily dependent on the amount of data trimmed and ramp model used. In Figure \ref{fig:Spitzer} we also denote the ramp that has the lowest Bayesian Information Criterion (BIC) at each trim level with a black circle around the data point. This further demonstrates that the 4.5 $\micron$ data sets are no longer beholden to the whims of the systematic removal with the best-fit ramp the same across all trim levels, while the 3.6 $\micron$ events best-fit ramp depends on the trim level. To address this, we select trimming that results in all ramps measuring the approximately the same depths based on the 3.6 $\micron$ events, and adopt the same trimming at 4.5 $\micron$ for consistency. We select 90 minutes of trimming for the transits, and 30 minutes of trimming for the eclipses. 

\begin{figure*}
    \centering
    \includegraphics[width = 0.24\textwidth]{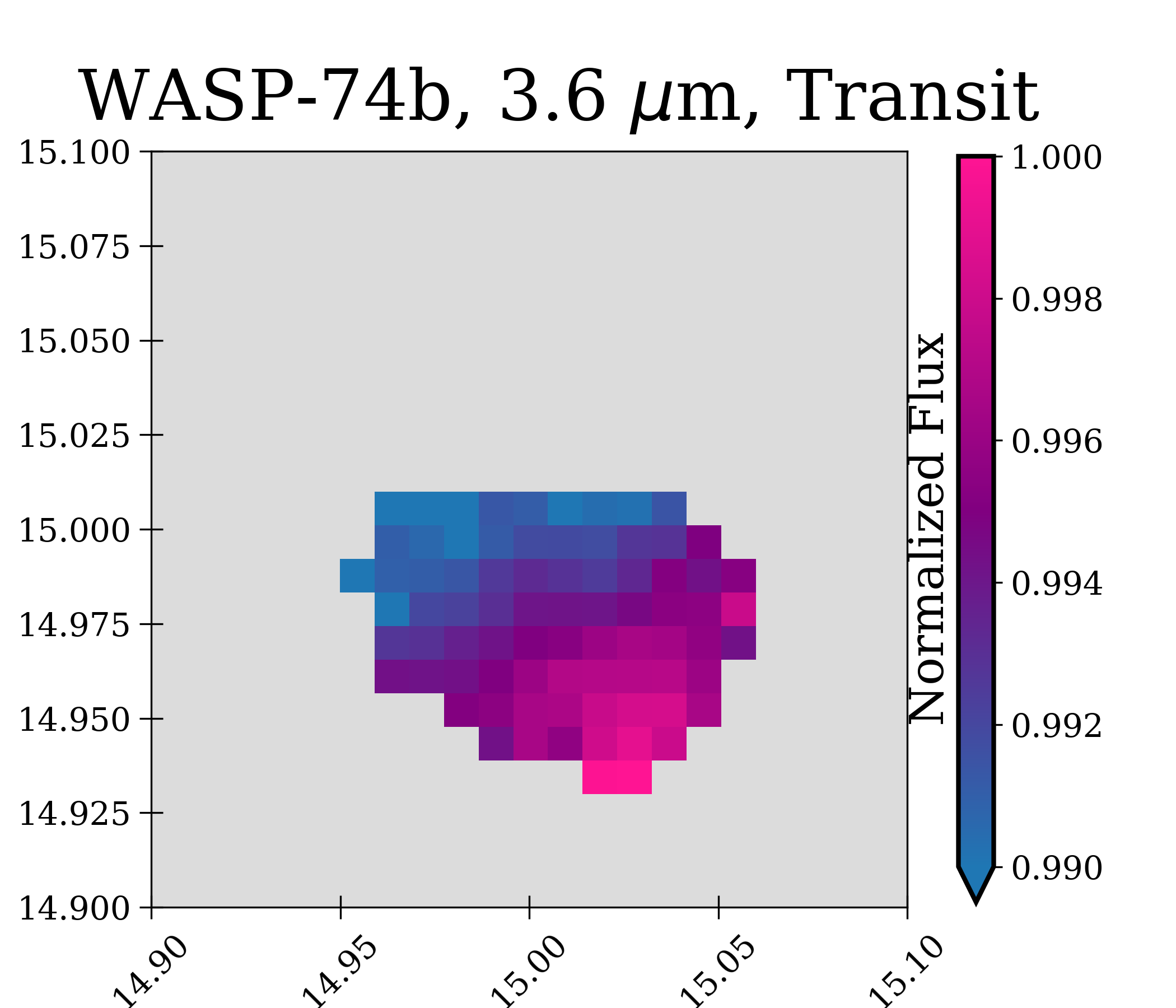}
    \includegraphics[width = 0.24\textwidth]{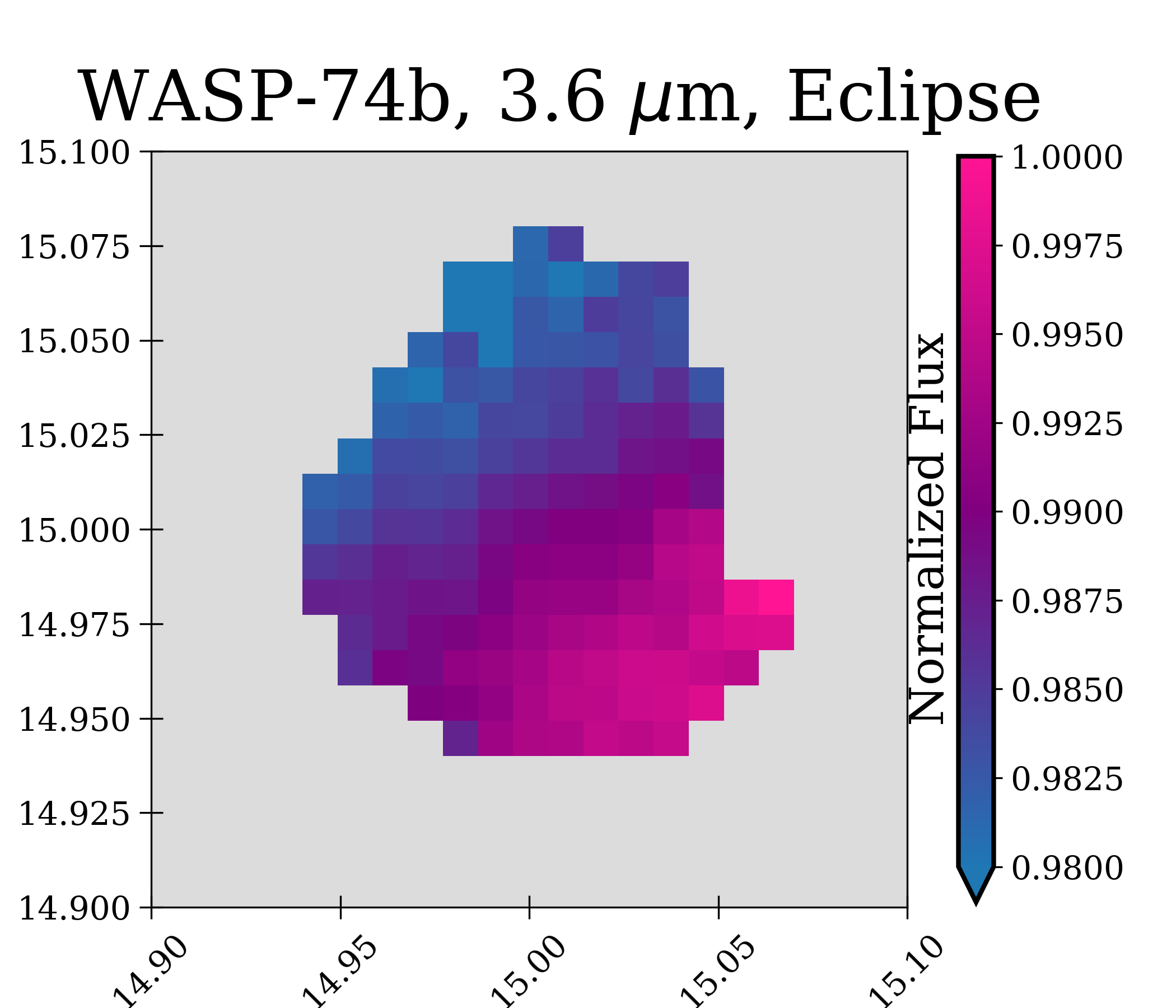}
    \includegraphics[width = 0.24\textwidth]{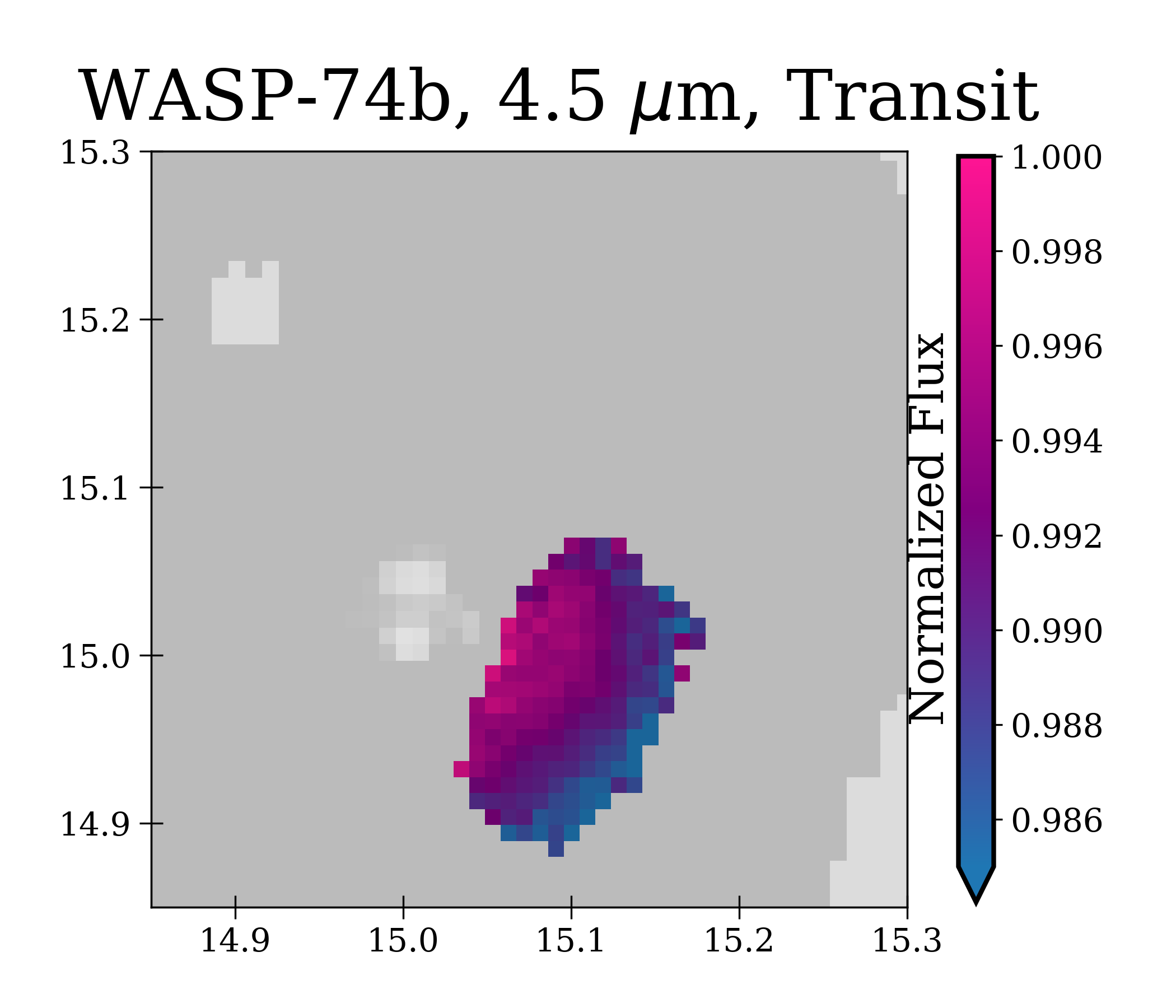}
    \includegraphics[width = 0.24\textwidth]{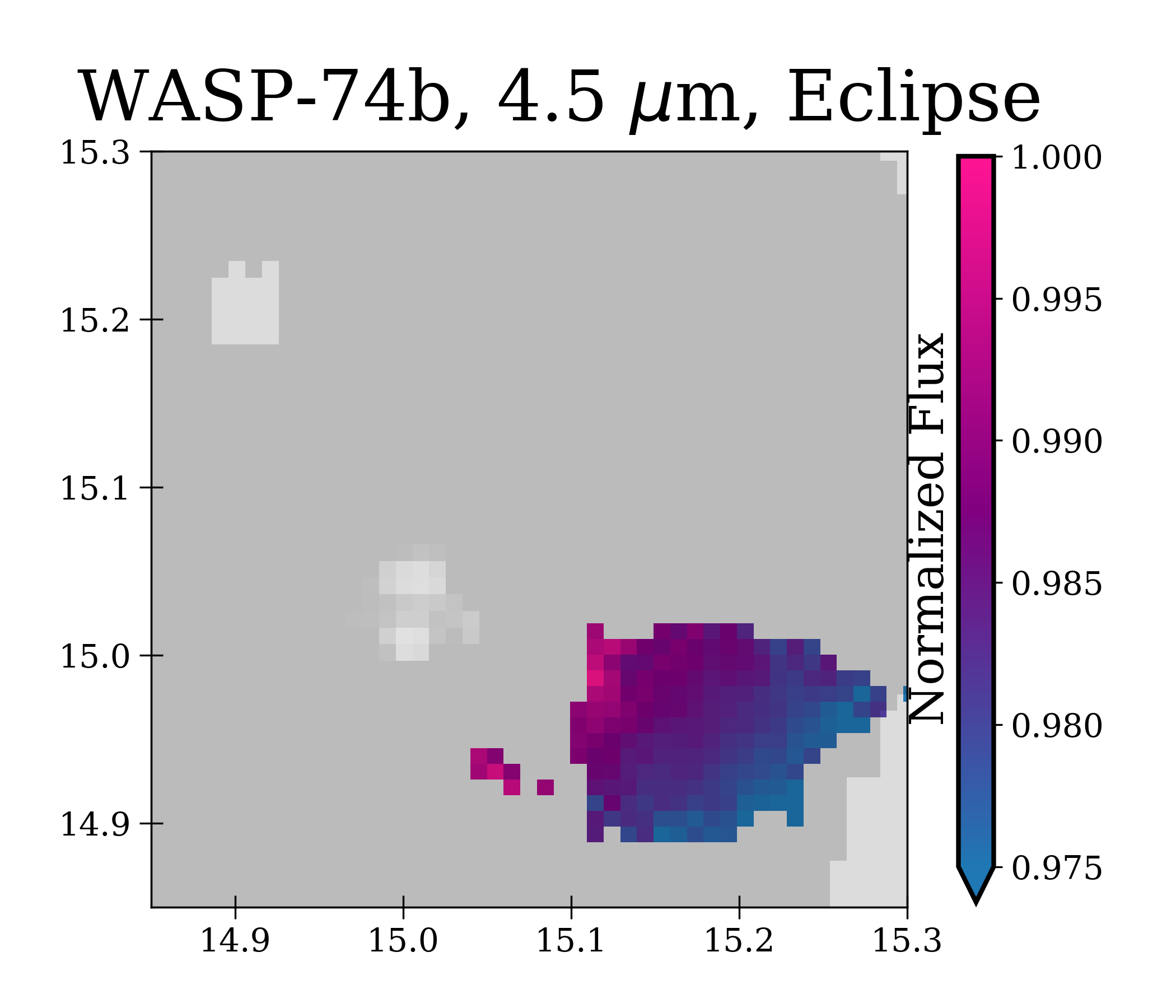}
    \caption{BLISS maps from best fits for all 4 Spitzer IRAC events as labeled. For the 4.5 $\micron$ maps, the darker grey regions denote where the fixed sensitivity map exists. Axis units are in sub-pixels. The BLISS map shown here is the subset of the fixed sensitivity map at the locations of the data centroids. It is not generated from this 4.5 micron dataset.}
    \label{fig:BLISS}
\end{figure*}

Our final fits use a linear ramp and 2$^{nd}$ order PRF-FWHM and standard BLISS map detrending at 3.6 $\micron$ with a 90 minute trim for the transit and a 30 minute trim for the eclipse. Our 4.5 $\micron$ fits use a linear ramp and fixed sensitivity map detrending for both events, with the same trimming as 3.6 $\micron$.

\section{Simultaneous ground-based photometric monitoring}

To search for the presence of magnetic activity in WASP-74, we performed
nightly photometry on the star during the 2018--2021 observing seasons with the Tennessee State University Celestron 14-inch (C14) automated imaging telescope (AIT) located at Fairborn Observatory in southern Arizona \citep[e.g.,][]{henry_transiting_2000, oswalt_future_2003}. The AIT is equipped with an SBIG STL-1001E CCD camera and uses a Cousins $R$ filter. Each observation consists of 3--5 consecutive exposures on WASP-74 and several surrounding comparison stars in the same CCD field of view. The individual frames are co-added and reduced to differential magnitudes in the sense WASP-74 minus the mean brightness of the comparison stars. Further details of our observing, reduction, and analysis techniques can be found in \cite{sing_hst_2015}. 

A total of 245 good nightly observations were collected between 2018 March 4 and 2021 June 18 (plus a few transit observations that we discarded). The observations are summarized in Table~\ref{monitor}. The standard deviation of the individual nightly observations from the mean of the complete data set is 0.00262~mag, which is close to the typical precision of a single observation with the C14.  The seasonal means agree to within 0.0007~mag. Thus, we confirm the findings of \cite{hellier_three_2015} that WASP-74 is magnetically inactive and photometrically constant. 

\begin{figure}
\centering
  \includegraphics[width=0.45\textwidth,keepaspectratio]{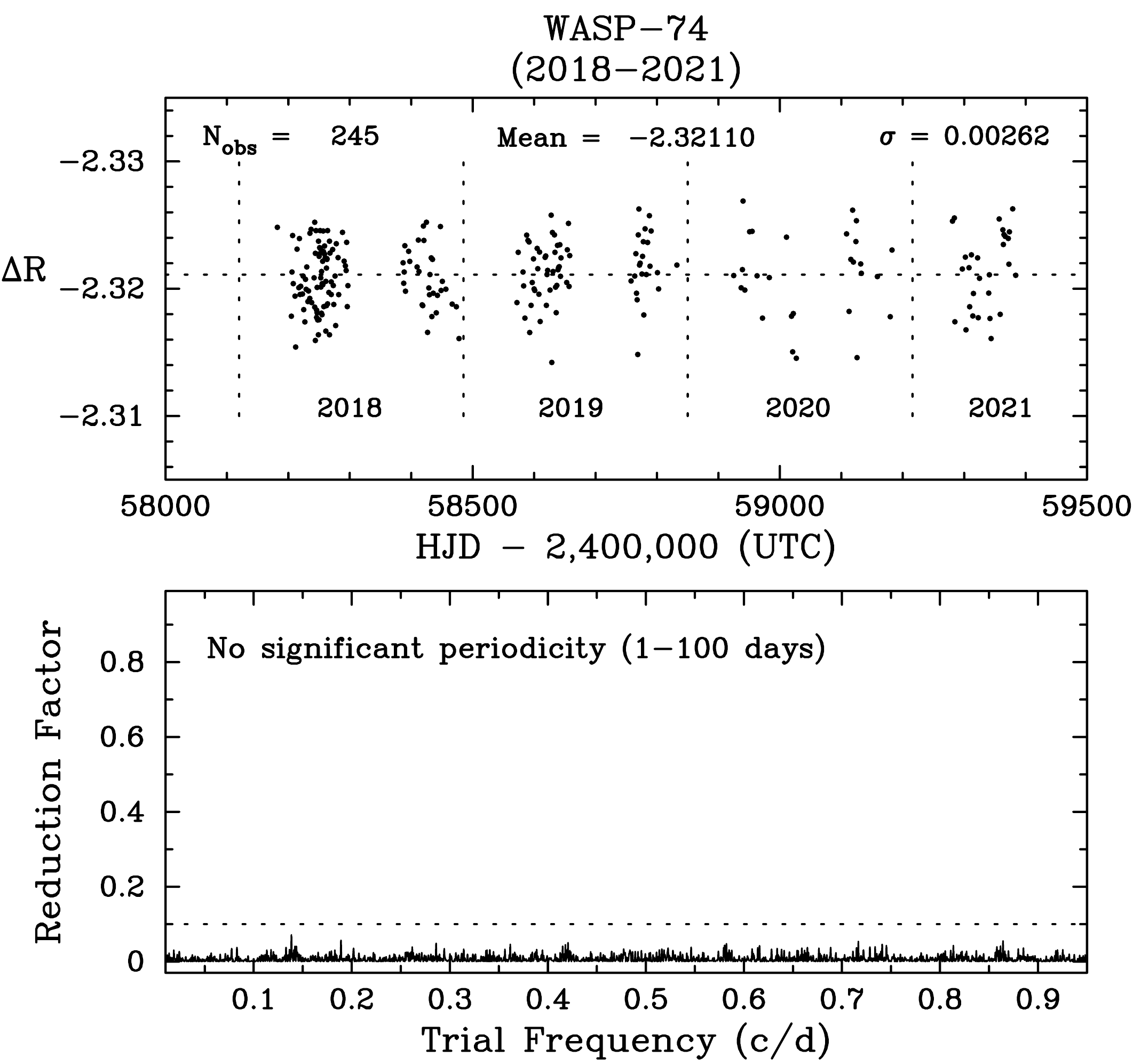}
  \caption{$Top$: The Cousins $R$ band photometry of WASP-74 from 2018--2021, acquired with the C14 automated imaging telescope (AIT) at Fairborn Observatory.  The star is constant on night-to-night and year-to-year time scales to the limit of our precision.  $Bottom$: Frequency spectrum of the complete data set shows no significant periodicity between 1 and 100 days. The horizontal dashed line in the bottom panel represents the noise limit below which reliable periods are not found in multi-color photometry \citep{henry_volume-limited_2011}.}
  \label{fig:ground_monitor}
\end{figure}

The photometric observations are plotted in the top panel of Figure~\ref{fig:ground_monitor} where the vertical dotted lines mark the beginning of each calendar year.  WASP-74 comes to opposition in late July; this means we lose the central part of each seasonal light curve because we must shut down our telescopes during the summer "monsoons" of southern Arizona (typically early July to early September). The summer gap is visible in the 2018, 2019, and 2020 observing seasons, but not in 2021 since the latest observations end at the shutdown. A frequency spectrum, based on least-squares fitting of sine curves, is shown in the bottom panel of Figure~\ref{fig:ground_monitor}. No significant periodicity is seen in the range 1--100 days.  Separate analyses of the individual observing seasons similarly showed no significant periodicities.

We phased the nightly observations to the radial-velocity period and computed a new least-squares sine fit.  The formal peak-to-peak amplitude of the photometry is $0.00093~\pm~0.00047$, consistent with zero to better than one mmag.  This confirms that radial velocity variations in WASP-74 are indeed due to planetary reflex motion and not line-profile variations due to spots \citep[e.g.,][]{queloz_no_2001, paulson_searching_2004}.

We did not detect any strong stellar variability with our three-years ground-based photometric monitoring campaign. Therefore any offsets in transit and eclipse depth between different observations are unlikely to be caused by changing in host star brightness from stellar activities such as varying starspots coverage.

\begin{deluxetable}{ccccc}
\tablecaption{SUMMARY OF C14 AIT PHOTOMETRY OF WASP-74}
\tablehead{
\colhead{Observing} & \colhead{} & \colhead{Date Range} &
\colhead{Sigma} & \colhead{Seasonal Mean} \\
\colhead{Season} & \colhead{$N_{obs}$} & \colhead{(HJD $-$ 2,400,000)} &
\colhead{(mag)} & \colhead{(mag)} \\
\colhead{(1)} & \colhead{(2)} & \colhead{(3)} &
\colhead{(4)} & \colhead{(5)}}
\startdata
 2018  & 119 & 58182--58477 & 0.00242 & $-$2.32081  \\
 2019  &  69 & 58572--58832 & 0.00243 & $-$2.32147  \\
 2020  &  28 & 58925--59182 & 0.00338 & $-$2.32104  \\
 2021  &  29 & 59281--59383 & 0.00302 & $-$2.32146  \\
\enddata
\end{deluxetable}
\label{monitor}

\section{Comparison with previous studies}

\begin{figure*}
\centering
  \includegraphics[width=\textwidth,keepaspectratio]{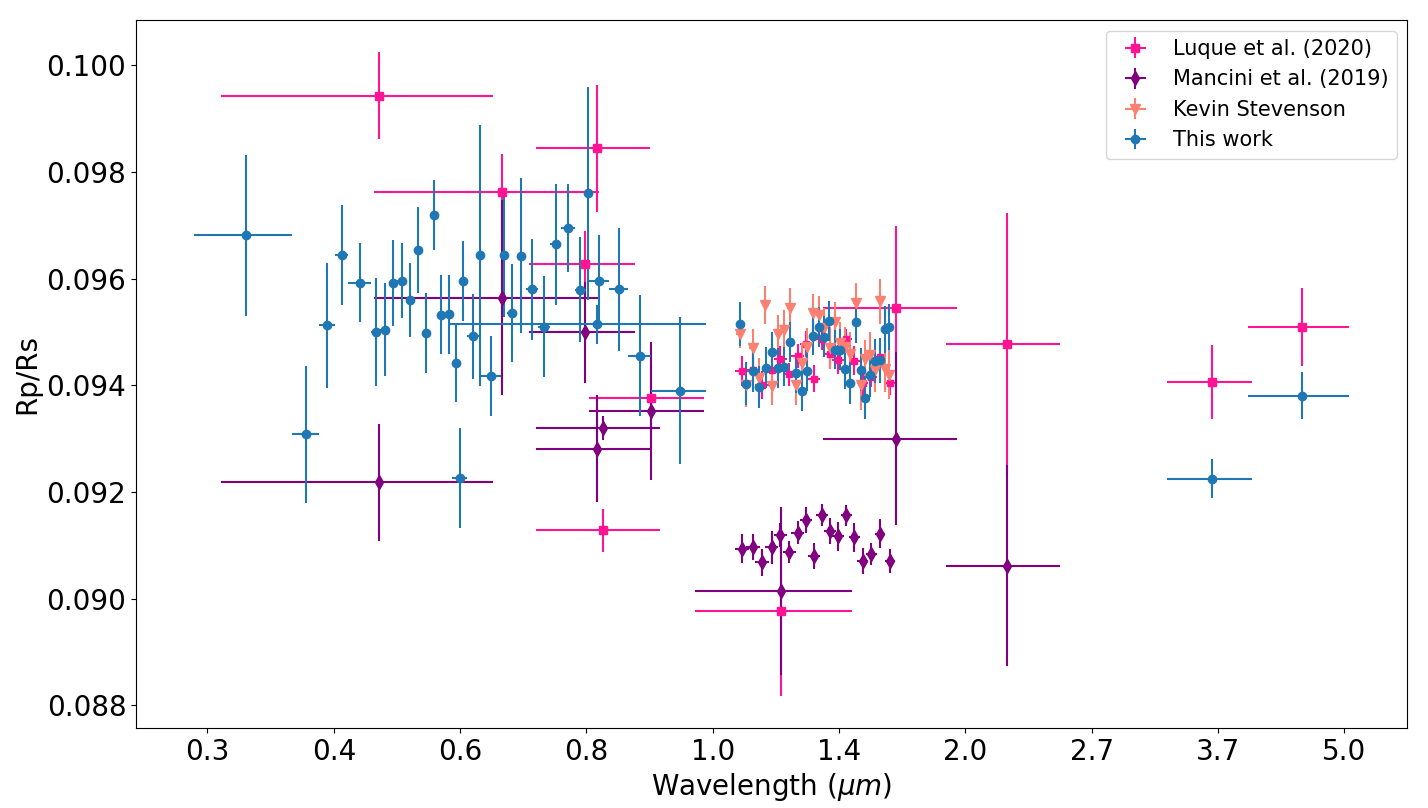}
  \caption{Comparison of transit spectra from this study to previous studies. We have fixed all orbital parameters to the exactly values as used in \citep{mancini_physical_2019}.}
  \label{fig:transit_spectra_compare}
\end{figure*}

We have compared our transmission spectrum with previous studies from \cite{luque_obliquity_2020, mancini_physical_2019} which include multiple ground-based photometric points with WFC3 G141 and Spitzer datasets (Fig. \ref{fig:transit_spectra_compare}). System parameters (Table. \ref{W74}) used in data reduction for this work were fixed to the same exact values used in \cite{mancini_physical_2019} to avoid any offsets from lightcurve fitting.

\begin{figure}
\centering
  \includegraphics[width=0.45\textwidth,keepaspectratio]{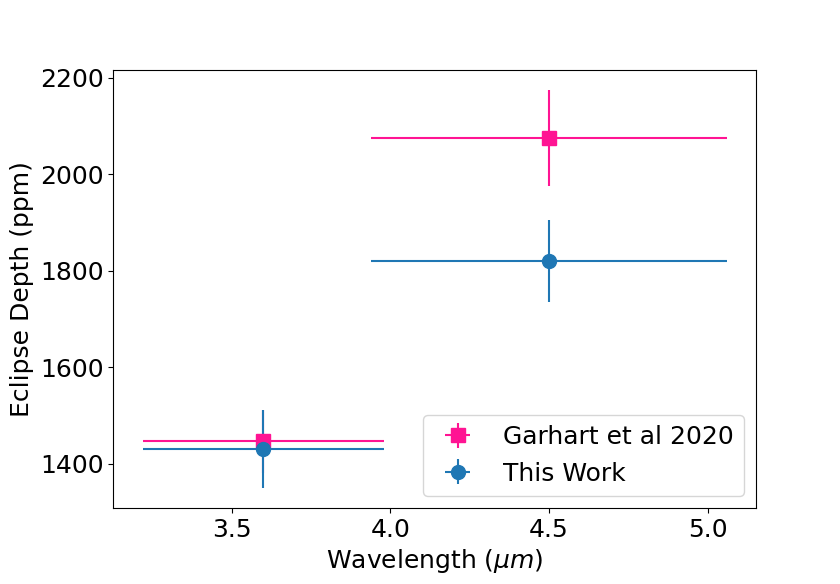}
  \caption{Comparison of Spitzer eclipse spectra from this study to \citep{garhart_statistical_2020}. We have fixed all orbital parameters to the exactly values as used in \cite{mancini_physical_2019}.}
  \label{fig:eclipse_spectra_compare}
\end{figure}

Our STIS spectrum sits between the ground-based photometry points in transit depth reported in \cite{mancini_physical_2019} and \cite{luque_obliquity_2020}. The tension between STIS and ground-based results could come from the different treatments used in reducing noisy ground-based lightcurves detailed in \cite{luque_obliquity_2020}. We find no evidence supporting neither strong TiO/VO absorption indicated by increased transit depth in the optical ($\sim$500 - 700nm) \citep{mancini_physical_2019} nor super-Rayleigh scattering \citep{luque_obliquity_2020}. 

The reduced WFC3 G141 transit spectrum has a large offset compared to \cite{mancini_physical_2019}. \cite{luque_obliquity_2020} took the exact same WFC3 G141 spectrum from \cite{mancini_physical_2019} and added an offset parameter in their PLATON retrieval. Our WFC3 G141 spectrum is highly consistent with the WFC3 offset version 2 reported in \cite{luque_obliquity_2020}. Kevin Stevenson (KS hereafter) also performed an independent analysis for the WFC3 G141 spectrum (Fig. \ref{fig:transit_spectra_compare}) with procedures detailed in \cite{stevenson_analyzing_2019, stevenson_thermal_2014} using the same exact orbital parameters used in \citep{mancini_physical_2019} and this work. Independently analyzed WFC3 G141 spectrum from KS is consistent with this work and WFC3 offset 2 version from \cite{luque_obliquity_2020} and also has a large offset compared to \cite{mancini_physical_2019}.

Our Spitzer points are lower than reported in \cite{luque_obliquity_2020} with channel 1 ($\sim3\sigma$ lower) having a larger offset than channel 2 ($\sim2\sigma$ lower). This could be due to the different choices of the minutes of data trimmed at the start of the observation and the ramp models. As shown in Figure \ref{fig:Spitzer}, we have extensively explored the different combinations of minutes to trim and ramp models. Our Spitzer spectrum represents transit depths from the best fitted lightcurves. 

We also compared our Spitzer eclipse spectrum with \cite{garhart_statistical_2020}. While the 3.6 $\mu m$ point is in agreement, our 4.5 $\mu m$ point is lower by $\sim$2.5$\sigma$. \cite{garhart_statistical_2020} used orbital parameters from the discovery paper \citep{hellier_three_2015} and we used the ones from \cite{mancini_physical_2019} to be consistent with the rest of this study.

\begin{center}
\begin{deluxetable}{cccc}
\tablecaption{PLATON transit retrieval results (full spectrum)}
\tablehead{\colhead{Parameter} & \colhead{Priors} & \colhead{Posteriors} & \colhead{Best fit}}
\startdata
\hline\hline 
T	&	$\mathcal{U}$(950, 1900)	&	$1288^{+282}_{-228}$	&	1557	\\
R$_{pl}$(Jup)	&	$\mathcal{U}$(1.1232, 1.404)	&	$1.33^{+0.02}_{-0.03}$	&	1.34	\\
log $f_{scatter}$	&	$\mathcal{U}$(0, 10)	&	$6.68^{+1.50}_{-1.89}$	&	3.97	\\
log(Z/$Z_{\Sun}$)	&	$\mathcal{U}$(-1, 1)	&	$0.03^{+0.61}_{-0.65}$	&	-0.37	\\
C/O ratio	&	$\mathcal{U}$(0.05, 2)	&	$0.85^{+0.72}_{-0.51}$	&	0.19	\\
log $P_{cloudtop}$ (Pa)	&	$\mathcal{U}$(0, 8)	&	$4.29^{+2.39}_{-2.39}$	&	7.08	\\
error multiple	&	$\mathcal{U}$(0.5, 5)	&	$1.25^{+0.11}_{-0.10}$	&	1.23	\\
scatter slope	&	$\mathcal{U}$(0, 10)	&	$2.81^{+0.94}_{-0.78}$	&	2.27	\\
\hline 
\enddata
\end{deluxetable}
\end{center}
\label{transit_priors}

\begin{center}
\begin{deluxetable}{cccc}
\tablecaption{PLATON transit retrieval results (without the 3.6 $\mu m$ point)}
\tablehead{\colhead{Parameter} & \colhead{Priors} & \colhead{Posteriors} & \colhead{Best fit}}
\startdata
\hline\hline 
T	&	$\mathcal{U}$(950, 1900)	&	$1413^{+333}_{-276}$	&	1458	\\
R$_{pl}$(Jup)	&	$\mathcal{U}$(1.1232, 1.404)	&	$1.35^{+0.02}_{-0.05}$	&	1.37	\\
log $f_{scatter}$	&	$\mathcal{U}$(0, 10)	&	$3.79^{+1.68}_{-1.32}$	&	2.82	\\
log(Z/$Z_{\Sun}$)	&	$\mathcal{U}$(-1, 1)	&	$-0.08^{+0.66}_{-0.59}$	&	-0.21	\\
C/O ratio	&	$\mathcal{U}$(0.05, 2)	&	$0.67^{+0.76}_{-0.40}$	&	0.81	\\
log $P_{cloudtop}$ (Pa)	&	$\mathcal{U}$(0, 8)	&	$4.33^{+2.27}_{-2.44}$	&	7.41	\\
error multiple	&	$\mathcal{U}$(0.5, 5)	&	$1.15^{+0.11}_{-0.09}$	&	1.07	\\
scatter slope	&	$\mathcal{U}$(0, 10)	&	$2.23^{+1.33}_{-0.90}$	&	1.56	\\
\hline 
\enddata
\end{deluxetable}
\end{center}
\label{transit_priors_no36}

\begin{center}
\begin{deluxetable}{cccc}
\tablecaption{PLATON eclipse retrieval results (full spectrum)}
\tablehead{\colhead{Parameter} & \colhead{Priors} & \colhead{Posteriors} & \colhead{Best fit}}
\startdata
\hline\hline 
log $f_{scatter}$	&	$\mathcal{U}$(0, 3)	&	$1.0^{+0.79}_{-0.64}$	&	1.16	\\
log(Z/$Z_{\Sun}$)	&	$\mathcal{U}$(-1, 3)	&	$-0.30^{+0.66}_{-0.46}$	&	-0.76	\\
C/O ratio	&	$\mathcal{U}$(0.1, 2)	&	$1.31^{+0.41}_{-0.26}$	&	0.90	\\
$\beta$	&	$\mathcal{U}$(0, 1.25)	&	$0.88^{+0.13}_{-0.16}$	&	0.46	\\
log $\kappa_{IR}$	&	$\mathcal{U}$(-4, -0.5)	&	$-1.80^{+0.44}_{-0.68}$	&	-1.20	\\
log ${\gamma}$	&	$\mathcal{U}$(-3, 0.5)	&	$-0.80^{+0.28}_{-0.37}$	&	-2.01	\\
\hline 
\enddata
\end{deluxetable}
\end{center}
\label{eclipse_priors}

\begin{center}
\begin{deluxetable}{cccc}
\tablecaption{PLATON eclipse retrieval results (without the 3.6 $\mu m$ point)}
\tablehead{\colhead{Parameter} & \colhead{Priors} & \colhead{Posteriors} & \colhead{Best fit}}
\startdata
\hline\hline 
log $f_{scatter}$	&	$\mathcal{U}$(0, 3)	&	$1.03^{+0.74}_{-0.62}$	&	0.86	\\
log(Z/$Z_{\Sun}$)	&	$\mathcal{U}$(-1, 3)	&	$-0.28^{+0.86}_{-0.49}$	&	0.03	\\
C/O ratio	&	$\mathcal{U}$(0.1, 2)	&	$0.92^{+0.30}_{-0.06}$	&	0.92	\\
$\beta$	&	$\mathcal{U}$(0, 1.25)	&	$0.86^{+0.16}_{-0.18}$	&	0.87	\\
log $\kappa_{IR}$	&	$\mathcal{U}$(-4, -0.5)	&	$-1.86^{+0.45}_{-0.70}$	&	-2.09	\\
log ${\gamma}$	&	$\mathcal{U}$(-3, 0.5)	&	$-0.88^{+0.33}_{-0.43}$	&	-0.86	\\
\hline 
\enddata
\end{deluxetable}
\end{center}
\label{eclipse_priors_no36}

\section{PLATON retrieval}
\subsection{Transit retrieval}
\begin{figure*}
\centering
  \includegraphics[width=\textwidth,keepaspectratio]{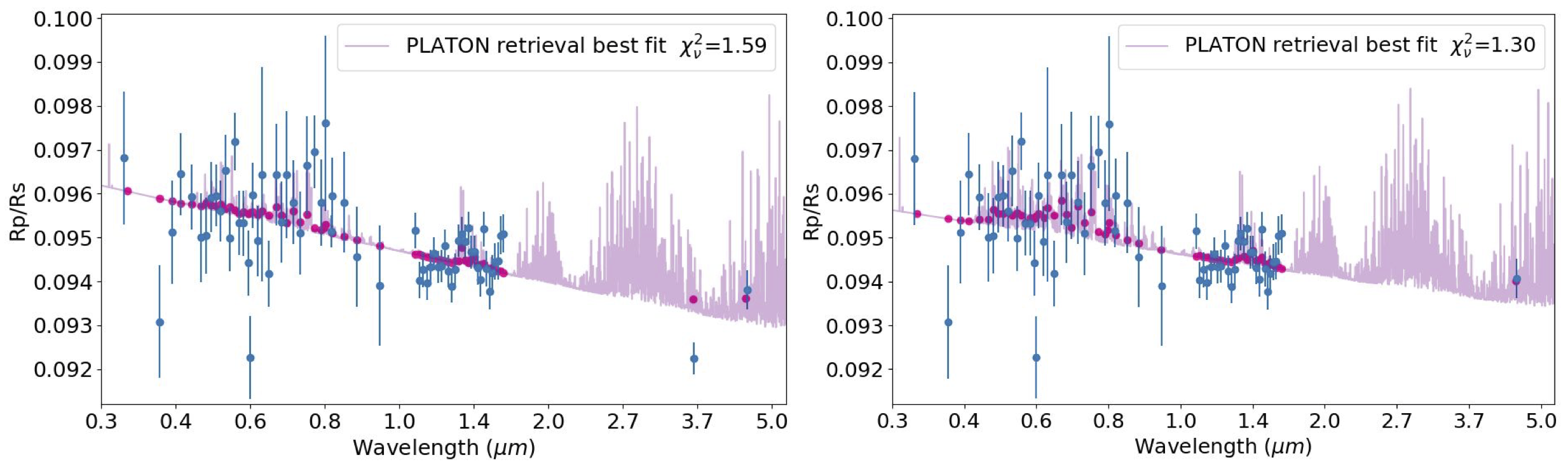}
  \caption{Best-fit PLATON retrieval of WASP-74b transit spectrum. Purple line is the full resolution model with pink dots for corresponding binned model. Left panel uses the full spectrum and the right panel excludes the 3.6 $\mu m$ point. We see strong scattering with muted water feature and no evidence for TiO absorption or super-Rayleigh scattering in both retrievals. However, the low 3.6 $\mu m$ point does drive up the cloud scattering factor.}
  \label{fig:transit_retrieval}
\end{figure*}

We performed a retrieval analysis with PLATON \citep{zhang_platon_2020} on both uniformly analyzed transit and eclipse spectra of WASP-74b. All retrievals were conducted with opacity sampling at R=1000 and 1000 live points. Our setup for the transit retrieval is shown in Table \ref{transit_priors} with seven free parameters assuming equilibrium chemistry and including condensation. The best-fit PLATON retrieval model with a $\chi^2_{\nu}$ of 1.59 is shown in Figure \ref{fig:transit_retrieval}. The model shows a very muted 1.4 $\mu m$ water feature accompanied with a strong scattering slope extending from the optical to infrared.

We found metallicity and C/O ratio consistent with solar values \ref{fig:transit_corner}. The retrieved limb averaged temperature from the transit spectrum based on 1D models is typically lower compared to the equilibrium temperature in hot Jupiters. This is because atmospheric circulation causes an always colder western limb to the eastern limb \citep{showman_atmospheric_2009} and 1D models retrieve a single uniform temperature for both limbs which could be biased towards lower temperature by several hundred degrees \citep{macdonald_why_2020}. In the colder western limb, aluminum and silicate based aerosols can form under these cooler environments which lead to strong scattering effect and muted molecular absorption features \citep{parmentier_transitions_2016, munoz_probing_2015}. 

We also performed a retrieval using the same priors but without the 3.6 $\mu m$ Spitzer point (Figure \ref{fig:transit_retrieval}). The Spitzer channel 1 transit depth can vary by several hundred ppm based on the trimming choice at the beginning of the lightcurve (Figure \ref{fig:Spitzer}). Excluding the 3.6 $\mu m$ point allows us to investigate if our retrieval result hinges on this one data point. The low transit depth of the 3.6 $\mu m$ point does drive up the retrieved log scattering factor (Table \ref{transit_priors} $\&$ \ref{transit_priors_no36}) from $3.79^{+1.68}_{-1.32}$ to $6.68^{+1.50}_{-1.89}$ as expected since a stronger scattering is needed to fit for the steeper slope. However, all other retrieved parameters stay consistent with or without the 3.6 $\mu m$ point.

\subsection{Eclipse retrieval}

\begin{figure*}
\centering
  \includegraphics[width=\textwidth,keepaspectratio]{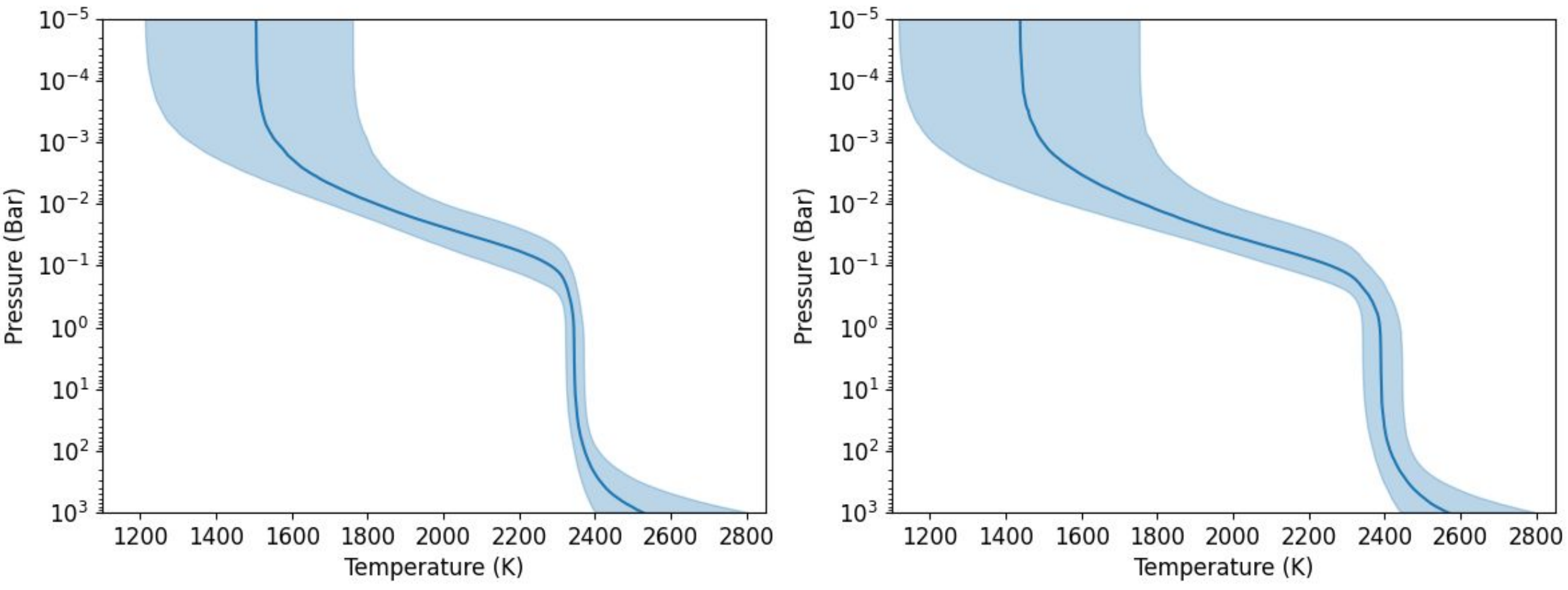}
  \caption{TP profile from PLATON eclipse retrieval with the solid line as the median values and shaded regions representing one sigma uncertainty. Left panel uses the full spectrum and the right panel excludes the 3.6 $\mu m$ point. The TP profiles are highly consistent showing the thermal structure is not sensitive to the one 3.6 $\mu m$ point.}
  \label{fig:eclipse_retrieval_TP}
\end{figure*}

For the eclipse retrieval we also assumed equilibrium chemistry with a total of six free parameters (Table \ref{eclipse_priors}) including three TP profile parameters \citep{line_systematic_2013}. The T$_{int}$ is set to 100K. The best-fit model with a $\chi^2_{\nu}$ of 1.47 is shown in Figure \ref{fig:eclipse_spectra_compare} and the TP profile is shown in Figure \ref{fig:eclipse_retrieval_TP} with dash lines indicating one sigma uncertainty. The WFC3/G141 emission spectrum is very close to blackbody-like with a potentially weak water absorption feature. The Spitzer 3.6 $\mu m$ point is $\sim3\sigma$ lower than a blackbody of 2260K assuming dayside heat redistribution which indicates a possible molecular absorption features. The strongest opacity source covered by the 3.6 $\mu m$ band is CH$_4$. Although large quantities of methane is unlikely to exist under a dayside temperature $>2000K$, some amount could be present in the higher altitude and cooler environment combined with a high C/O ratio atmosphere \citep{moses_chemical_2013}.   

\begin{figure*}
\centering
  \includegraphics[width=\textwidth,keepaspectratio]{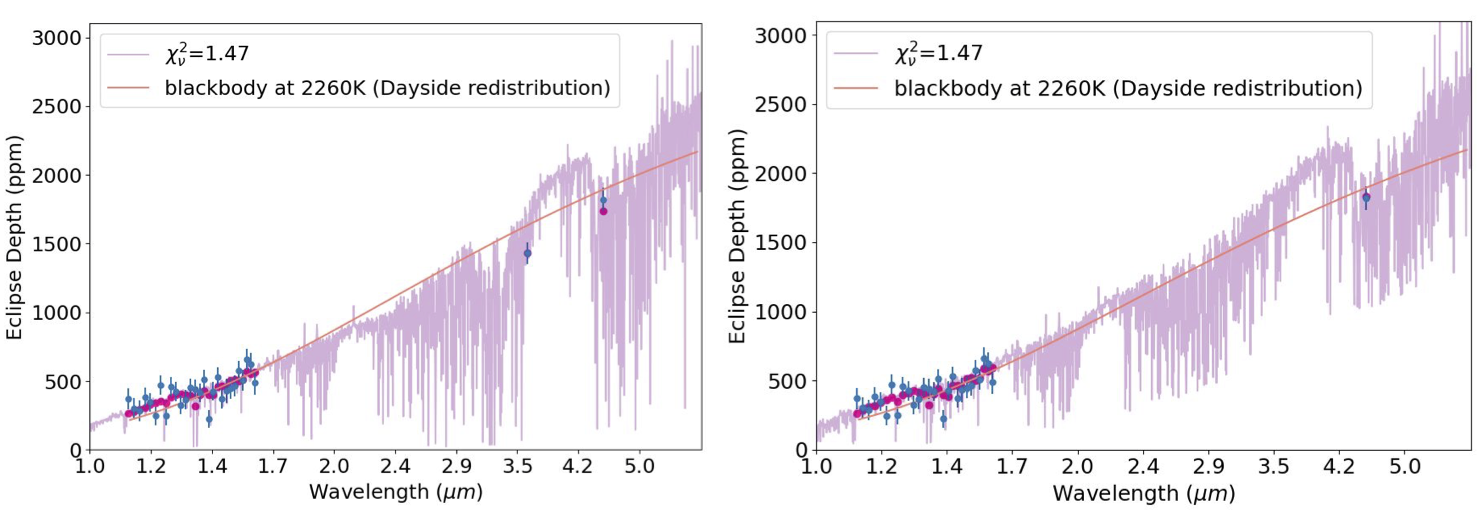}
  \caption{Best-fit PLATON retrieval of WASP-74b eclipse spectrum. Left panel uses the full spectrum and the right panel excludes the 3.6 $\mu m$ point. Purple line is the full resolution model with pink dots for corresponding binned model and overplotted with blackbody at 2260K in orange represent dayside only heat redistribution. In the full spectrum (left), we see a featureless blackbody-like WFC3/G141 spectrum with a methane absorption feature at 3.6 $\mu m$ band. Without the 3.6 $\mu m$ point (right), the methane absorption feature disappears and the retrieved C/O ratio drops significantly.}
  \label{fig:eclipse_retrieval}
\end{figure*}

The emission spectrum probes deeper into the atmosphere at around 1-5 bar and the retrieved decreasing TP profile at those pressure levels is consistent within one sigma compared to what we expect from dayside heat redistribution (2260K) on WASP-74b. The retrieved high C/O ratio of 1.31 is surprising since it would indicate very low water abundance with high abundance carbon bearing species such as CO and CH$_4$. We believe this high C/O ratio is driven by the featureless G141 spectrum combined with the low Spitzer 3.6$\mu m$ point. This combination led to a low water abundance solution which suppresses the water absorption feature one would expect in a decreasing TP profile while maintaining the methane absorption feature in the Spitzer channel 1 band. 

We believe three potential causes could have explained this unusually high C/O ratio indicated by the emission spectrum. One would be the reliability of Spitzer 3.6 $\mu m$ point which has been shown to be challenging to analyze due to the more complex instrument systematics compared to the 4.5 $\mu m$ channel \citep{may_introducing_2020}. A higher 3.6 $\mu m$ point with no methane absorption feature would indicate a more isothermal TP profile which would not need an extremely low water abundance to fit the featureless G141 spectrum. Another cause could be the present of an unknown opacity source in the G141 band masking out the water absorption feature. We know H- can have this effect \citep{lothringer_extremely_2018, arcangeli_climate_2019, parmentier_thermal_2018}, but it only exists in ultra-hot Jupiters (>2500K) and WASP-74b is not hot enough for H- to become a dominate opacity source. We have ran the retrievals with H- turned on and off and there are no changes on the retrieval results. High C/O ratio could also be due to planet formation beyond the snowline where the oxygen was locked up in the water ice while carbon was still free as CO in the gaseous phase \citep{madhusudhan_co_2012, oberg_effects_2011}. 

We also performed a retrieval on the emission spectrum without the 3.6 $\mu m$ point to see how much this one data point affects the retrieval result. The C/O ratio drops significantly (Table \ref{eclipse_priors} $\&$ \ref{eclipse_priors_no36}) from $1.31^{+0.41}_{-0.26}$ to $0.92^{+0.30}_{-0.06}$ showing the low 3.6 $\mu m$ point is driving up the methane abundance and thus the C/O ratio. All other parameters and the TP profile stayed consistent.

\section{Follow up observations with JWST}

The surprisingly high C/O ratio retrieval from the eclipse spectrum calls for JWST follow up observations with extended wavelength coverage. Eclipse observations with NIRSpec G235 and G395 would cover multiple infrared features from CH$_4$, H$_2$O CO, and CO$_2$ which would allow us to precisely constrain the C/O ratio. For the transit spectrum, NIRISS SOSS will be able to confirm the strong scattering slope which extends from the optical into the infrared and better characterize the aerosol properties. 

WASP-74b sits in an interesting transition parameter space between hot Jupiters and ultra-hot Jupiters where molecular features are not yet fully diminished by thermal dissociation and continuum opacity sources like H-, but high altitude aerosols can still linger to flatten the spectrum. Follow up studies of WASP-74b with JWST will better our understanding of hot Jupiters atmospheres in general and how their atmospheres change as they transition into ultra-hot Jupiters.

\section{Conclusion}

We observed 6 transits and 3 eclipses of the hot Jupiter WASP-74b with HST STIS/WFC3 and Spitzer. All datasets were uniformly analyzed with the same orbital parameters and limb darkening coefficients to ensure the consistence between different instruments. We have compared our transit spectrum with previous studies and we found no evidence for neither strong TiO/VO absorption in the optical as reported in \cite{mancini_physical_2019} nor super-Rayleigh scattering slope from \cite{luque_obliquity_2020}. Instead we found a muted water feature with strong aerosol scattering extending from the optical into the infrared. Both metallicity and C/O ratio are consistent to the solar values within one sigma. 

The eclipse retrieval results were more surprising with the preferred high C/O driven by possible CH$_4$ absorption feature in the Spitzer 3.6 $\mu m$ band. However, this result is highly sensitive to a single Spitzer data point and future JWST follow-up observations are needed for further investigations. 

\bibliography{references} 

\clearpage

\begin{appendix}

\begin{figure*}
\centering
  \includegraphics[width=\textwidth,keepaspectratio]{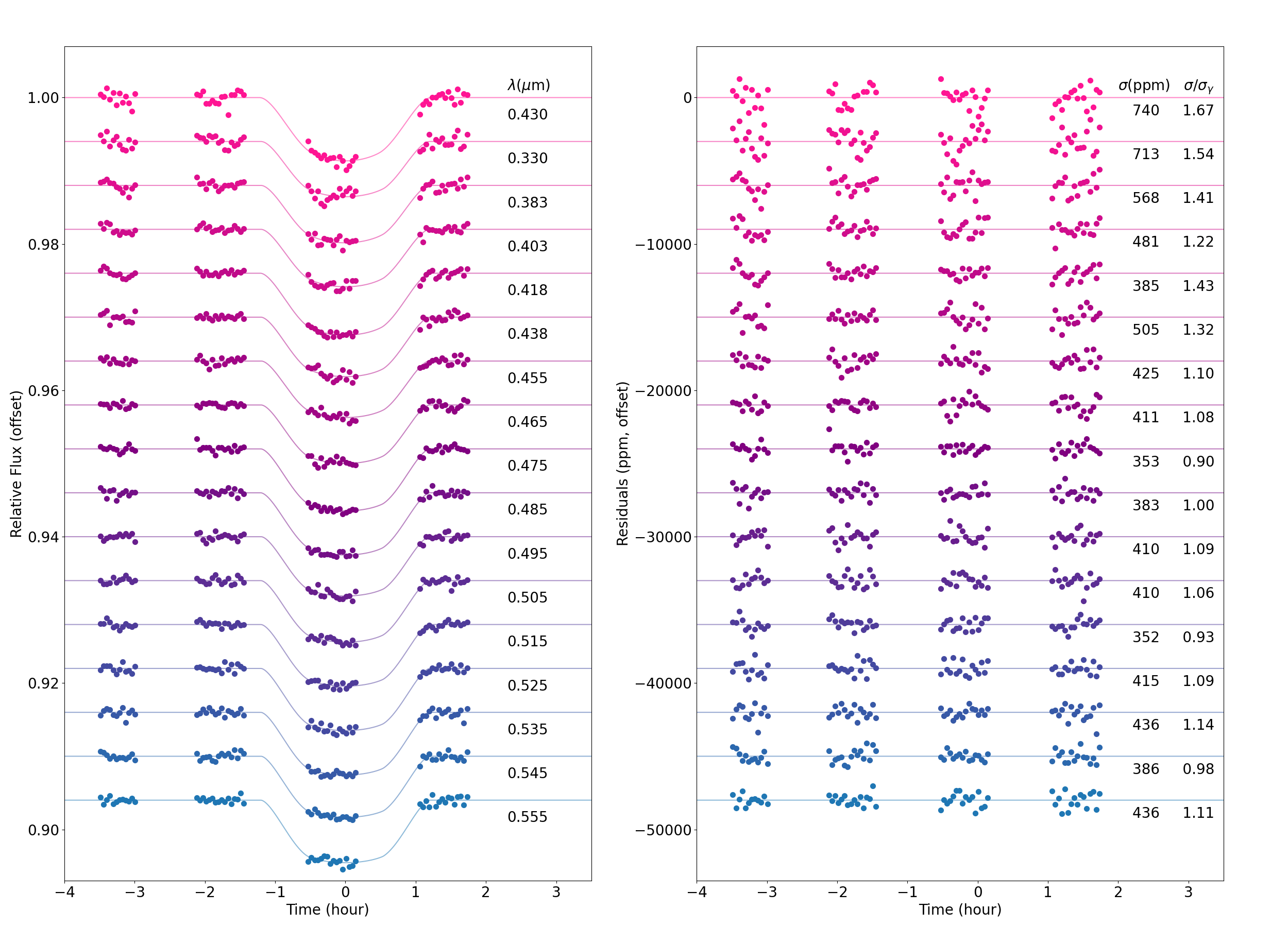}
  \caption{ HST STIS G430L visit 1 spectral bin transit lightcurves after systematics correction using jitter decorrelation (left) and corresponding residuals (right).}
  \label{fig:STIS_v76}
\end{figure*}

\begin{figure*}
\centering
  \includegraphics[width=\textwidth,keepaspectratio]{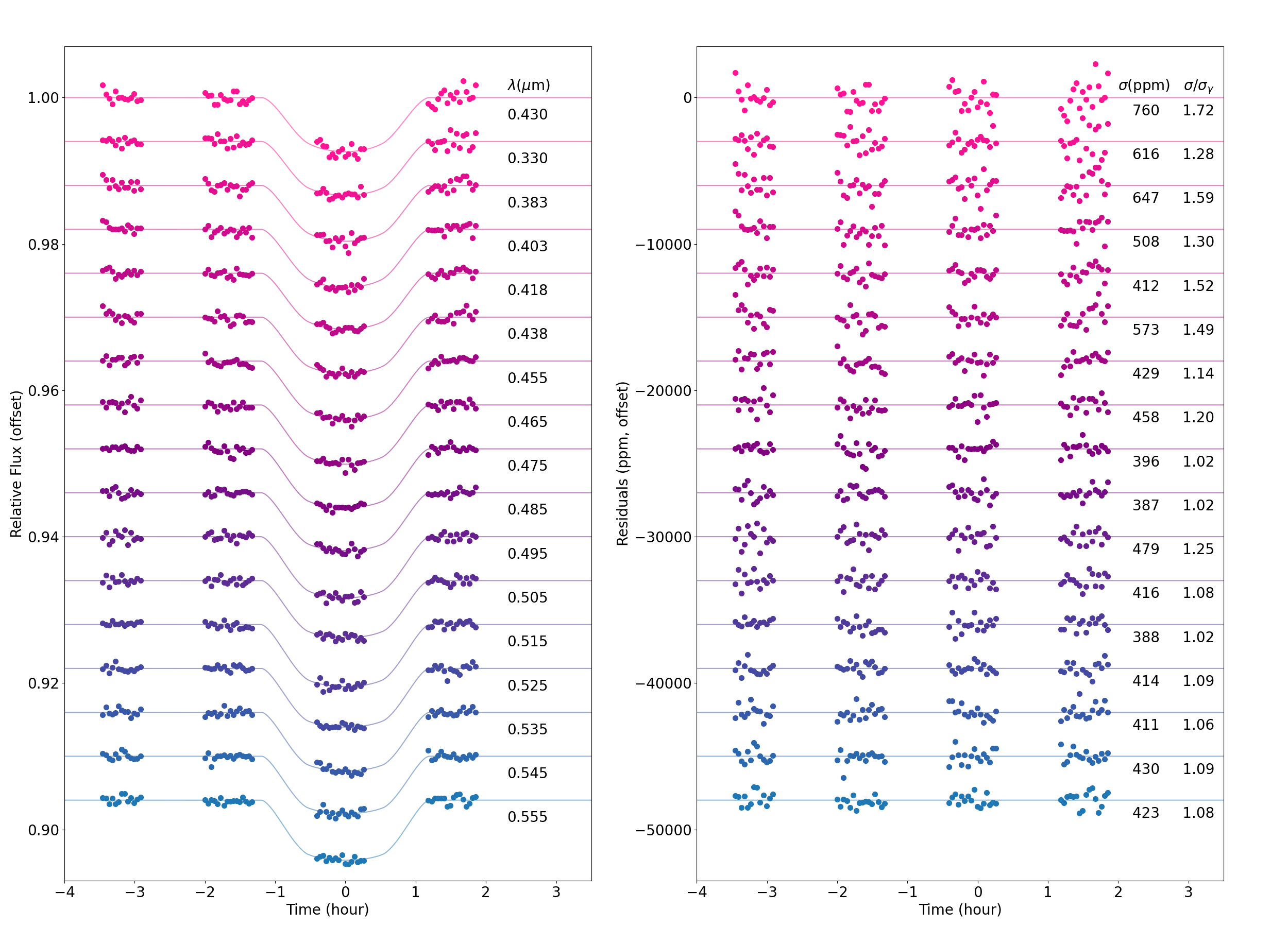}
  \caption{ HST STIS G430L visit 2 spectral bin transit lightcurves after systematics correction using jitter decorrelation (left) and corresponding residuals (right).}
  \label{fig:STIS_v77}
\end{figure*}

\begin{figure*}
\centering
  \includegraphics[width=\textwidth,keepaspectratio]{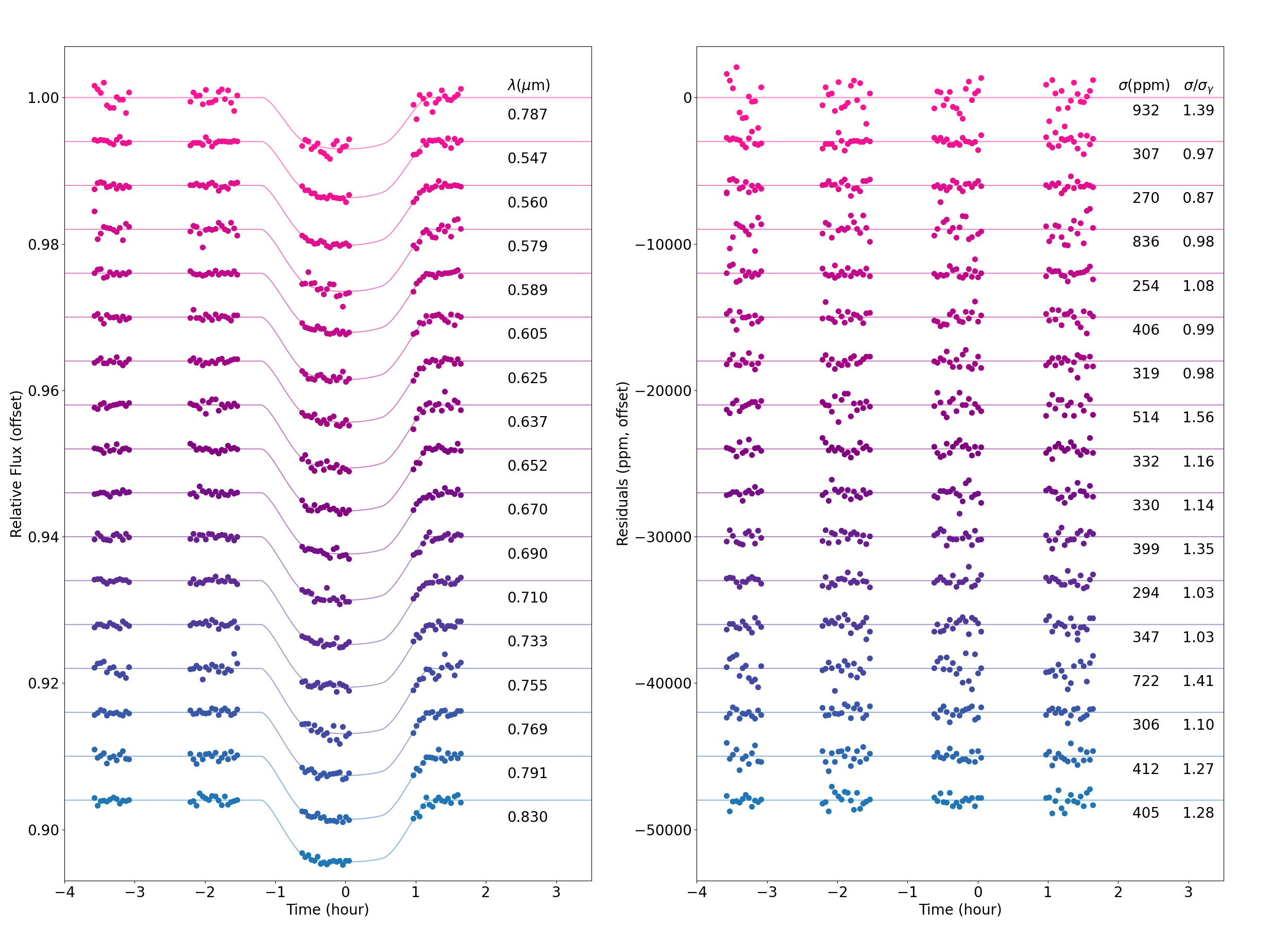}
  \caption{ HST STIS G750L spectral bin transit lightcurves after systematics correction using jitter decorrelation (left) and corresponding residuals (right).}
  \label{fig:STIS_v78}
\end{figure*}

\begin{figure*}
\centering
  \includegraphics[width=\textwidth,keepaspectratio]{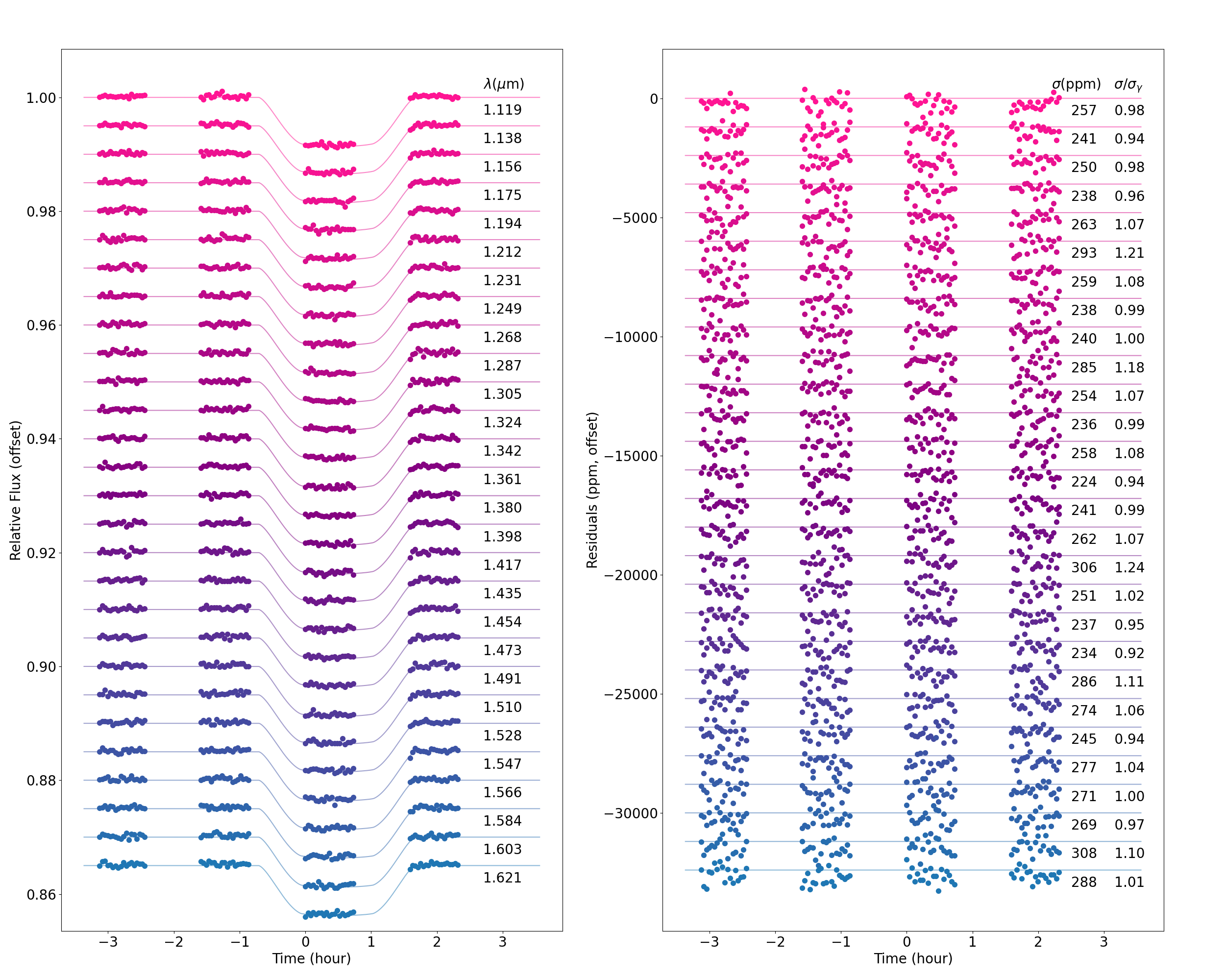}
  \caption{ HST WFC3 G141 spectral bin transit lightcurves after ramp-effect correction using RECTE (left) and corresponding residuals (right).}
  \label{fig:transit_wfc3_final}
\end{figure*}

\begin{figure*}
\centering
  \includegraphics[width=\textwidth,keepaspectratio]{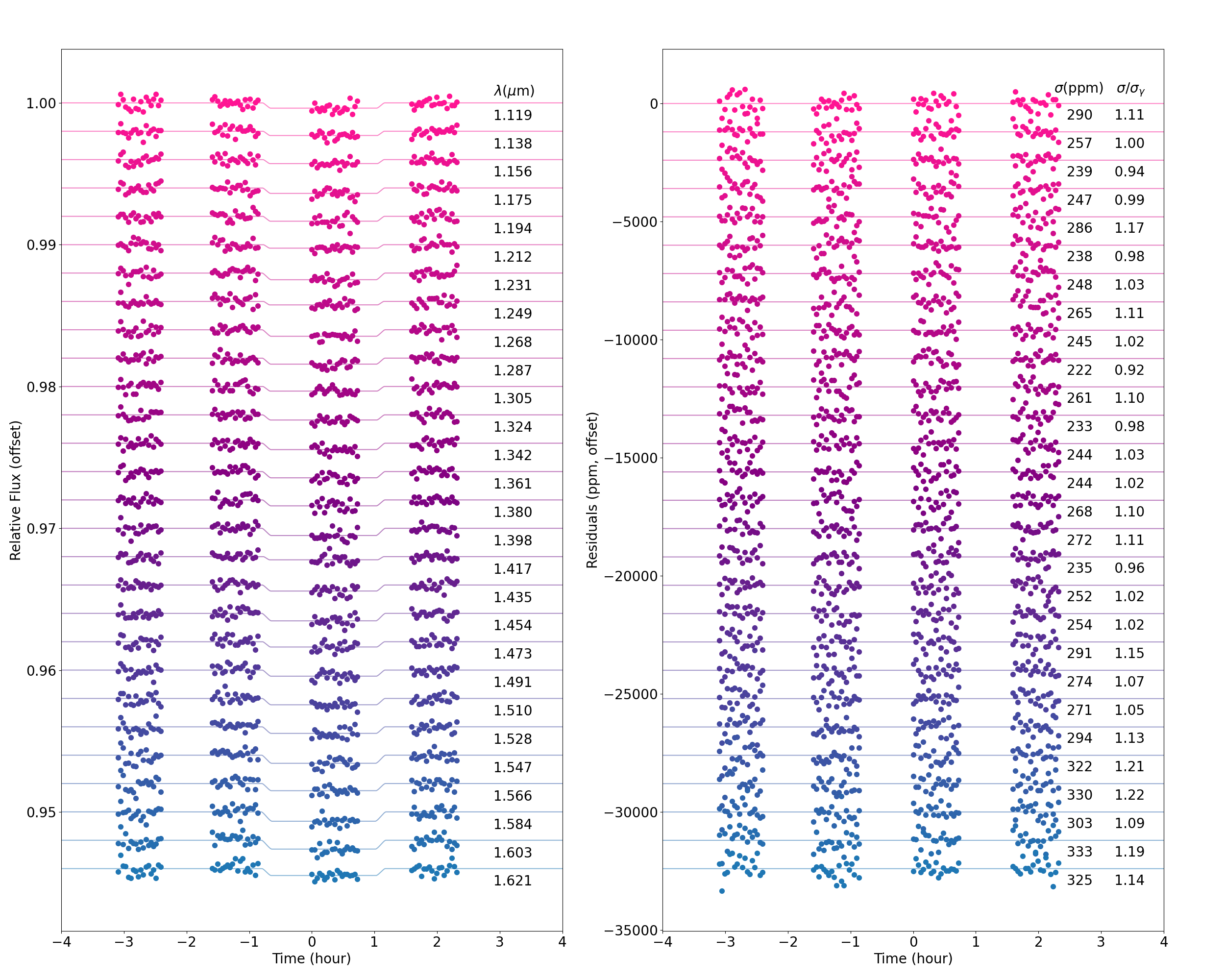}
  \caption{ HST WFC3 G141 spectral bin eclipse lightcurves after ramp-effect correction using RECTE (left) and corresponding residuals (right).}
  \label{fig:eclipse_wfc3_final}
\end{figure*}

\startlongtable
\begin{center}
\begin{deluxetable*}{cccc}
\tablecaption{\textbf{\large{WASP-74b transit spectrum}}}
\tablehead{\colhead{Wavelength midpoint ($\mu$m)} & \colhead{Bin width ($\mu$m)} & \colhead{Rp/Rs} & \colhead{Rp/Rs uncertainty}}
\startdata
\hline\hline 
0.330	&	0.0400	&	0.0968	&	0.001516	\\
0.382	&	0.0125	&	0.0931	&	0.001286	\\
0.403	&	0.0082	&	0.0951	&	0.001172	\\
0.418	&	0.0069	&	0.0964	&	0.000934	\\
0.438	&	0.0125	&	0.0959	&	0.000745	\\
0.455	&	0.0050	&	0.0950	&	0.001013	\\
0.465	&	0.0050	&	0.0950	&	0.000868	\\
0.475	&	0.0050	&	0.0959	&	0.000803	\\
0.485	&	0.0050	&	0.0960	&	0.000705	\\
0.495	&	0.0050	&	0.0956	&	0.000694	\\
0.505	&	0.0050	&	0.0965	&	0.000807	\\
0.515	&	0.0050	&	0.0950	&	0.000749	\\
0.525	&	0.0050	&	0.0972	&	0.000655	\\
0.535	&	0.0050	&	0.0953	&	0.000738	\\
0.545	&	0.0050	&	0.0953	&	0.000743	\\
0.555	&	0.0050	&	0.0944	&	0.000738	\\
0.560	&	0.0100	&	0.0923	&	0.000937	\\
0.565	&	0.0050	&	0.0960	&	0.000748	\\
0.579	&	0.0089	&	0.0949	&	0.000797	\\
0.589	&	0.0012	&	0.0964	&	0.002440	\\
0.605	&	0.0148	&	0.0942	&	0.000749	\\
0.625	&	0.0050	&	0.0964	&	0.001160	\\
0.637	&	0.0075	&	0.0954	&	0.000921	\\
0.652	&	0.0075	&	0.0964	&	0.001451	\\
0.670	&	0.0100	&	0.0958	&	0.000948	\\
0.690	&	0.0100	&	0.0951	&	0.000946	\\
0.710	&	0.0100	&	0.0966	&	0.001130	\\
0.733	&	0.0125	&	0.0969	&	0.000826	\\
0.755	&	0.0098	&	0.0958	&	0.000977	\\
0.769	&	0.0047	&	0.0976	&	0.001994	\\
0.788	&	0.2425	&	0.0951	&	0.000368	\\
0.791	&	0.0190	&	0.0960	&	0.000856	\\
0.830	&	0.0200	&	0.0958	&	0.001150	\\
0.875	&	0.0250	&	0.0946	&	0.001138	\\
0.965	&	0.0650	&	0.0939	&	0.001373	\\
1.119	&	0.0093	&	0.0952	&	0.000404	\\
1.138	&	0.0093	&	0.0940	&	0.000406	\\
1.156	&	0.0093	&	0.0943	&	0.000401	\\
1.175	&	0.0093	&	0.0940	&	0.000392	\\
1.194	&	0.0093	&	0.0943	&	0.000384	\\
1.212	&	0.0093	&	0.0946	&	0.000375	\\
1.231	&	0.0093	&	0.0943	&	0.000371	\\
1.249	&	0.0093	&	0.0943	&	0.000370	\\
1.268	&	0.0093	&	0.0948	&	0.000371	\\
1.287	&	0.0093	&	0.0942	&	0.000367	\\
1.305	&	0.0093	&	0.0939	&	0.000373	\\
1.324	&	0.0093	&	0.0943	&	0.000369	\\
1.342	&	0.0093	&	0.0949	&	0.000368	\\
1.361	&	0.0093	&	0.0951	&	0.000377	\\
1.380	&	0.0093	&	0.0949	&	0.000386	\\
1.398	&	0.0093	&	0.0952	&	0.000374	\\
1.417	&	0.0093	&	0.0947	&	0.000371	\\
1.435	&	0.0093	&	0.0947	&	0.000382	\\
1.454	&	0.0093	&	0.0943	&	0.000379	\\
1.473	&	0.0093	&	0.0940	&	0.000391	\\
1.491	&	0.0093	&	0.0952	&	0.000395	\\
1.510	&	0.0093	&	0.0943	&	0.000403	\\
1.528	&	0.0093	&	0.0938	&	0.000407	\\
1.547	&	0.0093	&	0.0942	&	0.000406	\\
1.566	&	0.0093	&	0.0944	&	0.000411	\\
1.584	&	0.0093	&	0.0945	&	0.000426	\\
1.603	&	0.0093	&	0.0951	&	0.000436	\\
1.621	&	0.0093	&	0.0951	&	0.000433	\\
3.600	&	0.3800	&	0.0922	&	0.000369	\\
4.500	&	0.5600	&	0.0938	&	0.000448	\\
\hline 
\enddata
\end{deluxetable*}
\end{center}
\label{transit_spectrum}

\startlongtable
\begin{center}
\begin{deluxetable*}{ccccc}
\tablecaption{\textbf{\large{Transit spectrum LD coefficients}}}\label{LD}
\tablehead{\colhead{Wavelength midpoint ($\mu$m)} & \colhead{C1} & \colhead{C2} & \colhead{C3} & \colhead{C4}}
\startdata
\hline\hline 
0.330	&	0.2705	&	0.3265	&	0.6655	&	-0.3603	\\
0.382	&	0.4105	&	0.0432	&	0.8133	&	-0.3888	\\
0.403	&	0.3063	&	0.3801	&	0.4784	&	-0.2826	\\
0.418	&	0.2976	&	0.4093	&	0.3791	&	-0.2082	\\
0.438	&	0.4077	&	0.3480	&	0.2256	&	-0.1518	\\
0.455	&	0.3666	&	0.2493	&	0.5645	&	-0.3223	\\
0.465	&	0.3696	&	0.3956	&	0.2662	&	-0.1798	\\
0.475	&	0.3562	&	0.5642	&	-0.0569	&	-0.0223	\\
0.485	&	0.4561	&	0.4177	&	0.0456	&	-0.0988	\\
0.495	&	0.4121	&	0.4911	&	-0.0624	&	-0.0201	\\
0.505	&	0.4444	&	0.3578	&	0.0620	&	-0.0501	\\
0.515	&	0.3952	&	0.3677	&	0.1715	&	-0.1388	\\
0.525	&	0.4643	&	0.3587	&	0.0349	&	-0.0625	\\
0.535	&	0.4823	&	0.3079	&	0.1146	&	-0.1165	\\
0.545	&	0.4839	&	0.3177	&	0.0417	&	-0.0598	\\
0.555	&	0.5090	&	0.2242	&	0.1408	&	-0.0968	\\
0.560	&	0.5141	&	0.2312	&	0.1016	&	-0.0735	\\
0.565	&	0.5190	&	0.2377	&	0.0652	&	-0.0518	\\
0.579	&	0.5348	&	0.2233	&	0.0450	&	-0.0415	\\
0.589	&	0.5811	&	-0.0467	&	0.5043	&	-0.2957	\\
0.605	&	0.5625	&	0.1408	&	0.1077	&	-0.0683	\\
0.625	&	0.5922	&	0.0536	&	0.1836	&	-0.1021	\\
0.637	&	0.5855	&	0.0663	&	0.1626	&	-0.0940	\\
0.652	&	0.6241	&	0.0128	&	0.1621	&	-0.0985	\\
0.670	&	0.6058	&	0.0120	&	0.1747	&	-0.0921	\\
0.690	&	0.6179	&	-0.0367	&	0.2073	&	-0.1001	\\
0.710	&	0.6250	&	-0.0672	&	0.2241	&	-0.1041	\\
0.733	&	0.6208	&	-0.0718	&	0.2233	&	-0.1084	\\
0.755	&	0.6350	&	-0.1154	&	0.2477	&	-0.1146	\\
0.769	&	0.6357	&	-0.1366	&	0.2639	&	-0.1211	\\
0.788	&	0.6053	&	-0.0234	&	0.1975	&	-0.0987	\\
0.791	&	0.6397	&	-0.1459	&	0.2567	&	-0.1149	\\
0.830	&	0.6343	&	-0.1428	&	0.2275	&	-0.1000	\\
0.875	&	0.6555	&	-0.2478	&	0.3210	&	-0.1341	\\
0.965	&	0.6439	&	-0.2223	&	0.2747	&	-0.1175	\\
1.119	&	0.6370	&	-0.2190	&	0.1870	&	-0.0668	\\
1.138	&	0.6340	&	-0.2280	&	0.1960	&	-0.0718	\\
1.156	&	0.6280	&	-0.1970	&	0.1500	&	-0.0514	\\
1.175	&	0.6390	&	-0.2200	&	0.1660	&	-0.0594	\\
1.194	&	0.6250	&	-0.1740	&	0.1070	&	-0.0324	\\
1.212	&	0.6280	&	-0.1790	&	0.0984	&	-0.0258	\\
1.231	&	0.6290	&	-0.1670	&	0.0776	&	-0.0201	\\
1.249	&	0.6350	&	-0.1770	&	0.0841	&	-0.0242	\\
1.268	&	0.6390	&	-0.1690	&	0.0600	&	-0.0120	\\
1.287	&	0.6680	&	-0.2030	&	0.0294	&	0.0052	\\
1.305	&	0.6460	&	-0.1590	&	0.0220	&	0.0075	\\
1.324	&	0.6520	&	-0.1700	&	0.0268	&	0.0075	\\
1.342	&	0.6630	&	-0.1770	&	0.0152	&	0.0147	\\
1.361	&	0.6660	&	-0.1820	&	0.0128	&	0.0171	\\
1.380	&	0.6910	&	-0.2130	&	0.0192	&	0.0190	\\
1.398	&	0.7010	&	-0.2280	&	0.0178	&	0.0241	\\
1.417	&	0.7210	&	-0.2570	&	0.0378	&	0.0165	\\
1.435	&	0.7290	&	-0.2800	&	0.0469	&	0.0187	\\
1.454	&	0.7500	&	-0.3260	&	0.0930	&	-0.0039	\\
1.473	&	0.7880	&	-0.3770	&	0.1120	&	-0.0026	\\
1.491	&	0.7960	&	-0.4150	&	0.1440	&	-0.0102	\\
1.510	&	0.8280	&	-0.4930	&	0.2160	&	-0.0380	\\
1.528	&	0.8600	&	-0.5370	&	0.2280	&	-0.0310	\\
1.547	&	0.8810	&	-0.6060	&	0.2970	&	-0.0573	\\
1.566	&	0.8890	&	-0.6450	&	0.3380	&	-0.0694	\\
1.584	&	0.8270	&	-0.5490	&	0.2740	&	-0.0540	\\
1.603	&	0.9480	&	-0.8340	&	0.5450	&	-0.1550	\\
1.621	&	0.9440	&	-0.7890	&	0.4750	&	-0.1210	\\
3.600	&	0.4958	&	-0.4478	&	0.3558	&	-0.1170	\\
4.500	&	0.4255	&	-0.3651	&	0.2867	&	-0.0945	\\
\hline 
\enddata
\end{deluxetable*}
\end{center}

\startlongtable
\begin{center}
\begin{deluxetable*}{cccc}
\tablecaption{\textbf{\large{WASP-74b eclipse spectrum}}}
\tablehead{\colhead{Wavelength midpoint ($\mu$m)} & \colhead{Bin width ($\mu$m)} & \colhead{Occultation Depth (ppm)} & \colhead{Uncertainty (ppm)}}
\startdata
\hline\hline 
1.1193	&	0.0093	&	372	&	76	\\
1.1379	&	0.0093	&	302	&	73	\\
1.1564	&	0.0093	&	284	&	74	\\
1.1751	&	0.0093	&	385	&	70	\\
1.1937	&	0.0093	&	346	&	68	\\
1.2123	&	0.0093	&	246	&	70	\\
1.2309	&	0.0093	&	473	&	69	\\
1.2494	&	0.0093	&	247	&	67	\\
1.2681	&	0.0093	&	460	&	68	\\
1.2867	&	0.0093	&	426	&	70	\\
1.3053	&	0.0093	&	326	&	67	\\
1.3238	&	0.0093	&	365	&	69	\\
1.3424	&	0.0093	&	452	&	68	\\
1.3611	&	0.0093	&	442	&	73	\\
1.3797	&	0.0093	&	418	&	67	\\
1.3982	&	0.0093	&	513	&	69	\\
1.4168	&	0.0093	&	224	&	63	\\
1.4354	&	0.0093	&	427	&	68	\\
1.4541	&	0.0093	&	529	&	71	\\
1.4727	&	0.0093	&	373	&	71	\\
1.4912	&	0.0093	&	431	&	87	\\
1.5098	&	0.0093	&	448	&	74	\\
1.5284	&	0.0093	&	465	&	75	\\
1.5471	&	0.0093	&	574	&	75	\\
1.5656	&	0.0093	&	506	&	75	\\
1.5842	&	0.0093	&	659	&	79	\\
1.6028	&	0.0093	&	624	&	79	\\
1.6214	&	0.0093	&	489	&	85	\\
3.6000	&	0.3800	&	1430	&	81	\\
4.5000	&	0.5600	&	1820	&	85	\\
\hline 
\enddata
\end{deluxetable*}
\end{center}
\label{eclipse_spectrum}

\begin{figure*}
\centering
  \includegraphics[width=\textwidth,keepaspectratio]{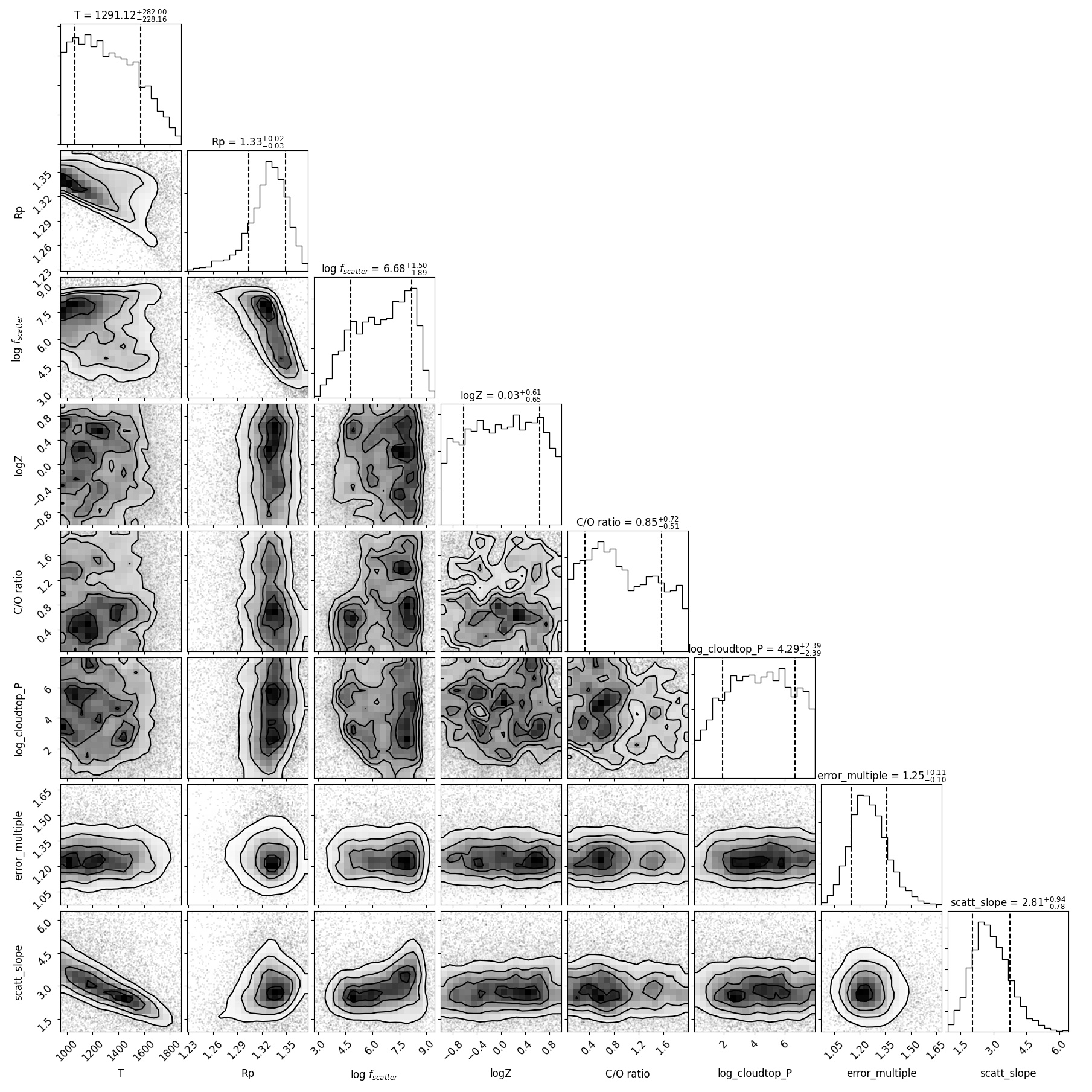}
  \caption{Posterior distribution of PLATON retrieval on the transit spectrum.}
  \label{fig:transit_corner}
\end{figure*}

\begin{figure*}
\centering
  \includegraphics[width=\textwidth,keepaspectratio]{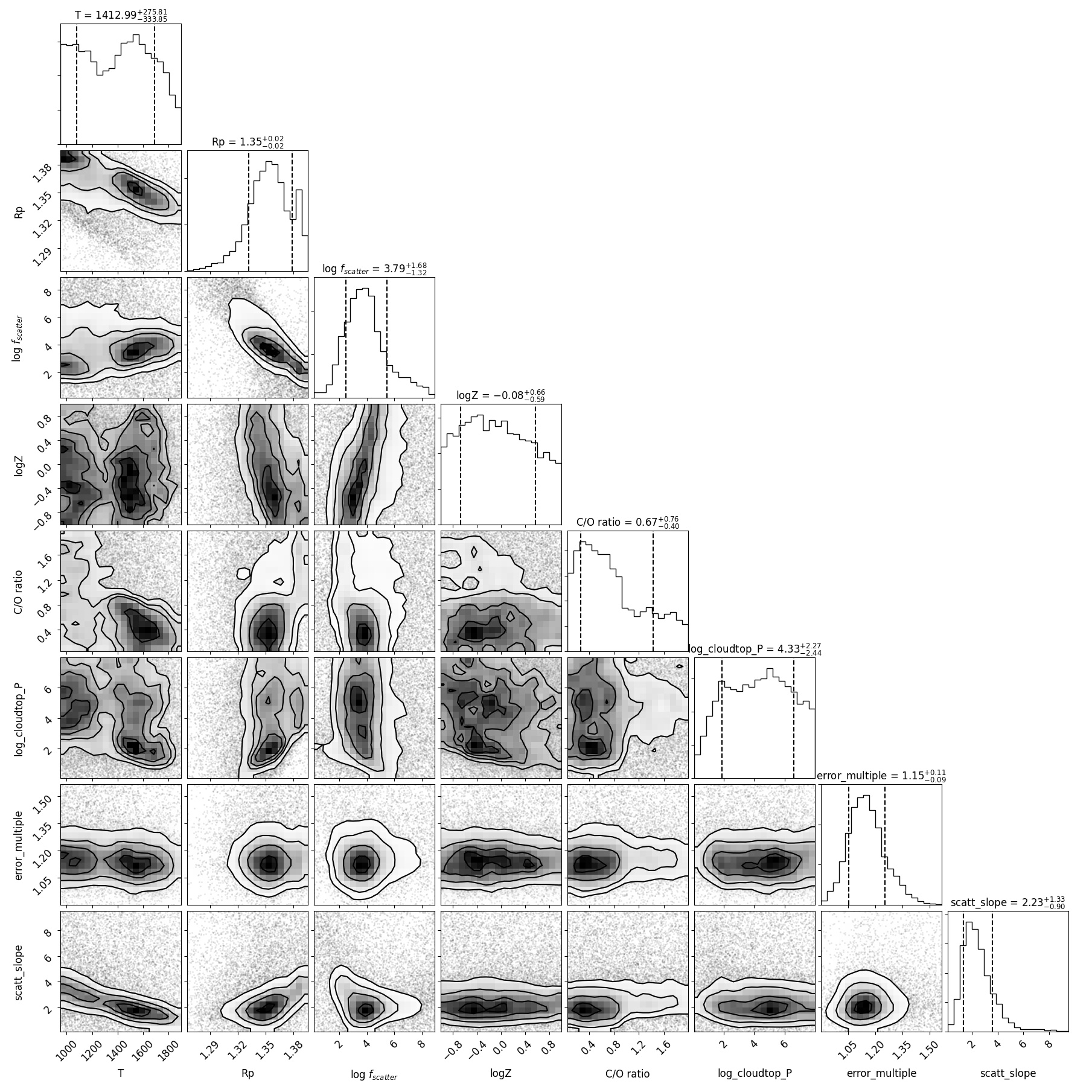}
  \caption{Posterior distribution of PLATON retrieval on the transit spectrum without the Spitzer 3.6 $\mu m$ point.}
  \label{fig:transit_corner_no36}
\end{figure*}

\begin{figure*}
\centering
  \includegraphics[width=\textwidth,keepaspectratio]{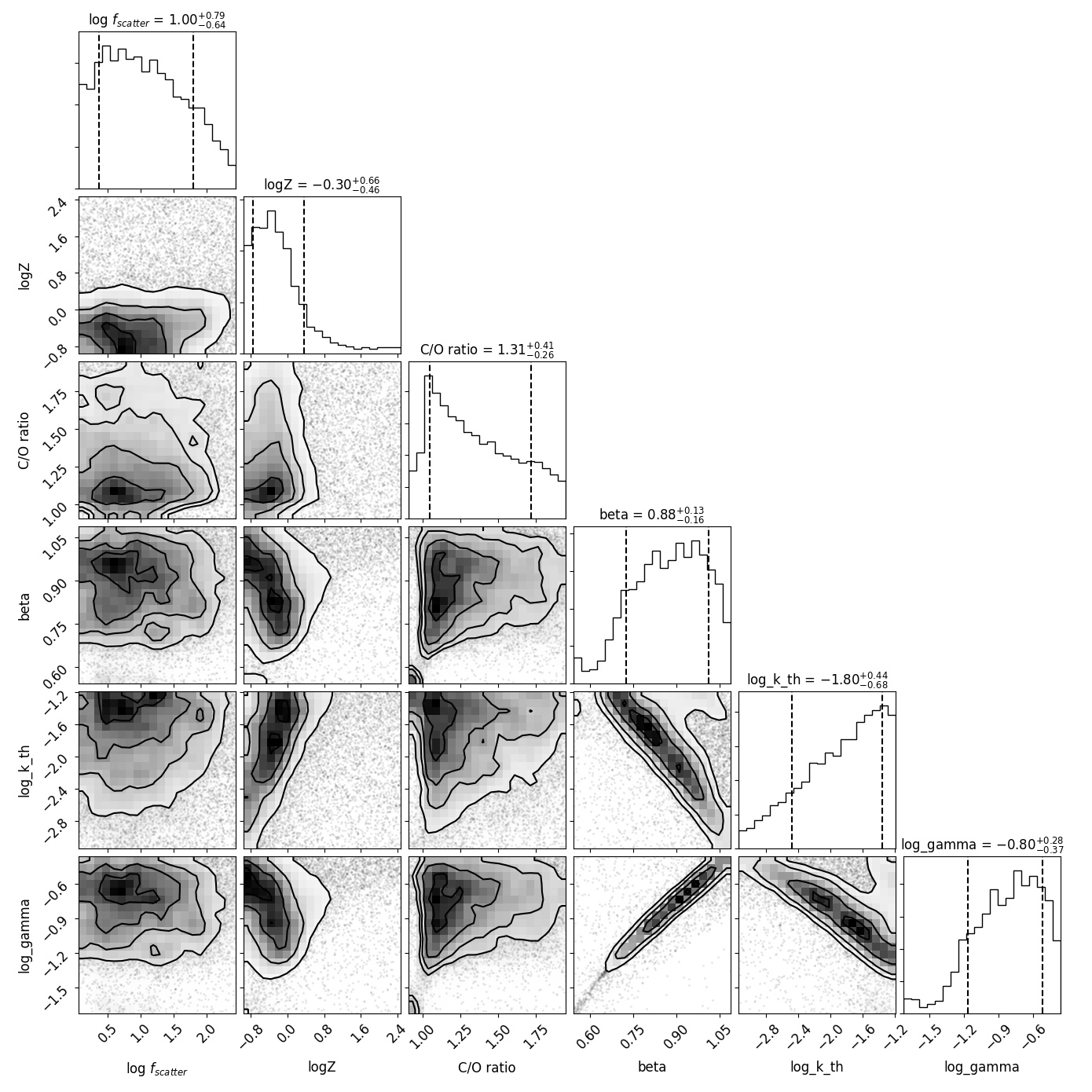}
  \caption{Posterior distribution of PLATON retrieval on the eclipse spectrum.}
  \label{fig:eclipse_corner}
\end{figure*}

\begin{figure*}
\centering
  \includegraphics[width=\textwidth,keepaspectratio]{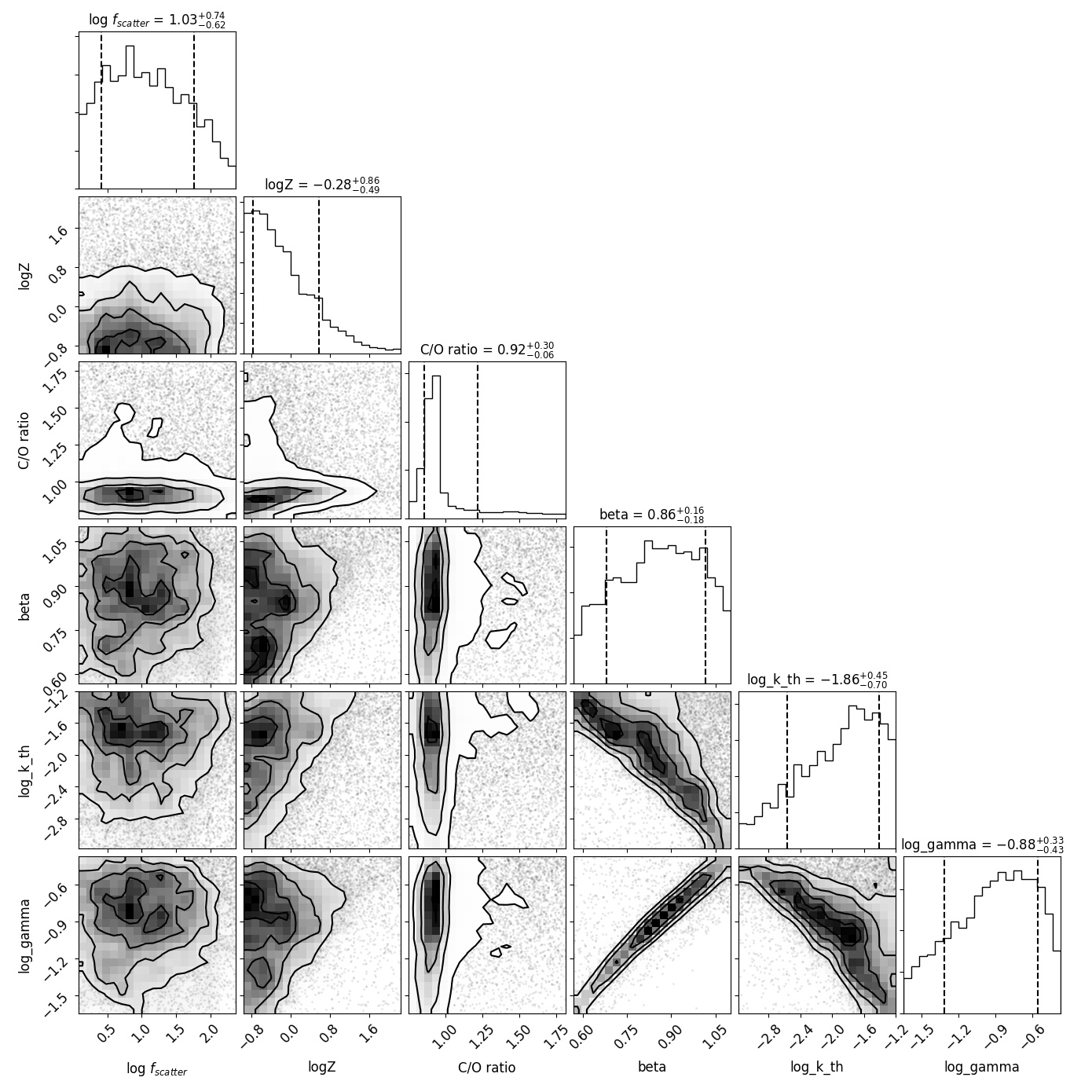}
  \caption{Posterior distribution of PLATON retrieval on the eclipse spectrum without the Spitzer 3.6 $\mu m$ point.}
  \label{fig:eclipse_corner_no36}
\end{figure*}

\end{appendix}

\end{document}